\documentclass[11pt]{article}
\usepackage[utf8]{inputenc}
\usepackage{siunitx}
\usepackage[margin=0.9in]{geometry}
\usepackage[english]{babel}
\usepackage{amsmath,amsfonts,amssymb}
\usepackage{graphicx}
\usepackage{authblk}
\usepackage[superscript]{cite}
\usepackage[explicit]{titlesec}
\usepackage{titletoc}
\usepackage{url}
\addto\captionsenglish{}

\title{Structure and magnetic properties of epitaxial CaFe\textsubscript{2}O\textsubscript{4} thin films}
\author[1,*]{Silvia Damerio}
\author[1]{Pavan Nukala}
\author[2]{Jean Juraszek}
\author[3]{Pim Reith}
\author[3]{Hans Hilgenkamp}
\author[1,4,*]{Beatriz Noheda}
\affil[1]{Zernike Institute for Advanced Materials, University of Groningen, 9747 AG Groningen, The Netherlands}
\affil[2]{Normandie Univ, UNIROUEN, INSA Rouen, CNRS, GPM, 76000 Rouen, France}
\affil[3]{Faculty of Science and Technology and MESA+ Institute for Nanotechnology, University of Twente, Enschede, The Netherlands}
\affil[4]{Groningen Cognitive Systems and Materials Center (CogniGron), 9747 AG Groningen, The Netherlands}
\affil[*]{email: s.damerio@rug.nl, b.noheda@rug.nl}
\date{}
\begin{document}
\maketitle
%%%%%%%%%%%%%%%%%%%%%%%%%%%%%%%%%%%%%%%%%%%%%%%%%%%%%%%%%%%%%%
\begin{abstract}
$CaFe_2O_4$ is a highly anisotropic antiferromagnet reported to display two spin arrangements with up-up-down-down (phase A) and up-down-up-down (phase B) configurations. The relative stability of these phases is ruled by the competing ferromagnetic and antiferromagnetic interactions between $Fe^{3+}$ spins arranged in two different environments, but a complete understanding of the magnetic structure of this material does not exist yet. In this study we investigate epitaxial $CaFe_2O_4$ thin films grown on $TiO_2$ (110) substrates by means of Pulsed Laser Deposition (PLD). Structural characterization reveals the coexistence of two out-of-plane crystal orientations and the formation of three in-plane oriented domains. The magnetic properties of the films, investigated macroscopically as well as locally, including highly sensitive Mössbauer spectroscopy, reveal the presence of just one order parameter showing long-range ordering below $T= 185$ K and the critical nature of the transition. In addition, a non-zero in-plane magnetization is found, consistent with the presence of uncompensated spins at phase or domain boundaries, as proposed for bulk samples.
\end{abstract}
%%%%%%%%%%%%%%%%%%%%%%%%%%%%%%%%%%%%%%%%%%%%%%%%%%%%%%%%
\section*{Introduction}
$CaFe_2O_4$ is an oxide semiconductor that, unlike most of the other ferrites with the same unit formula, does not have the Spinel structure\cite{BroesevanGroenou1969MagnetismFerrites} and, instead, crystallizes in a orthorhombic prototype structure with space group $Pnma$ and lattice parameters a=9.230\si{\angstrom} , b=3.024\si{\angstrom} and c=10.705\si{\angstrom} \cite{Decker1957TheFerrite, Galuskina2014HarmuniteIsrael}.\\
An extensive literature focuses on the catalytic activity of $CaFe_2O_4$ nanoparticles \cite{Lal2019RietveldNanoparticles, Khanna2013Size-dependentNanoparticles} and heterostructures \cite{Borse2012Ti-dopant-enhancedIrradiation,Cao2013PhotoelectrochemicalPhotoelectrodes,Ida2010PreparationWater, Kim2009FabricationPhotocatalysis}, with particular attention to its application as photo-cathode in $H_2$ generation and water splitting reactions. On the other hand, single crystals of this material are only moderately investigated \cite{Watanabe1967NeutronCaFe2O4, Yamamoto1968MossbauerCaFe2O4, Merlini2010Fe3+Pressure, Das2016Self-AdjustedCaFe2O4, Gandhi2017MagnetocrystallineCrystal, Stock2016SolitaryCaFe2O4,Stock2017OrphanCaFe2O4} and reports of epitaxial growth of $CaFe_2O_4$ thin films are almost absent \cite{Nishiyama2017HighlyDeposition}.\\
Since the first studies \cite{Watanabe1967NeutronCaFe2O4,Yamamoto1968MossbauerCaFe2O4}, the unusual magnetic structure of $CaFe_2O_4$ has been subject to debate and to date it has not yet been completely understood \cite{Das2018FerrimagneticCaFe2O4delta}. Recently, renewed interest in the topic has arisen following the neutron diffraction studies of Stock \textit{et al.}\cite{Stock2016SolitaryCaFe2O4,Stock2017OrphanCaFe2O4} on $CaFe_2O_4$ single crystals.\\

\begin{figure}[htb]
\begin{centering}
\includegraphics[width=.95\linewidth]{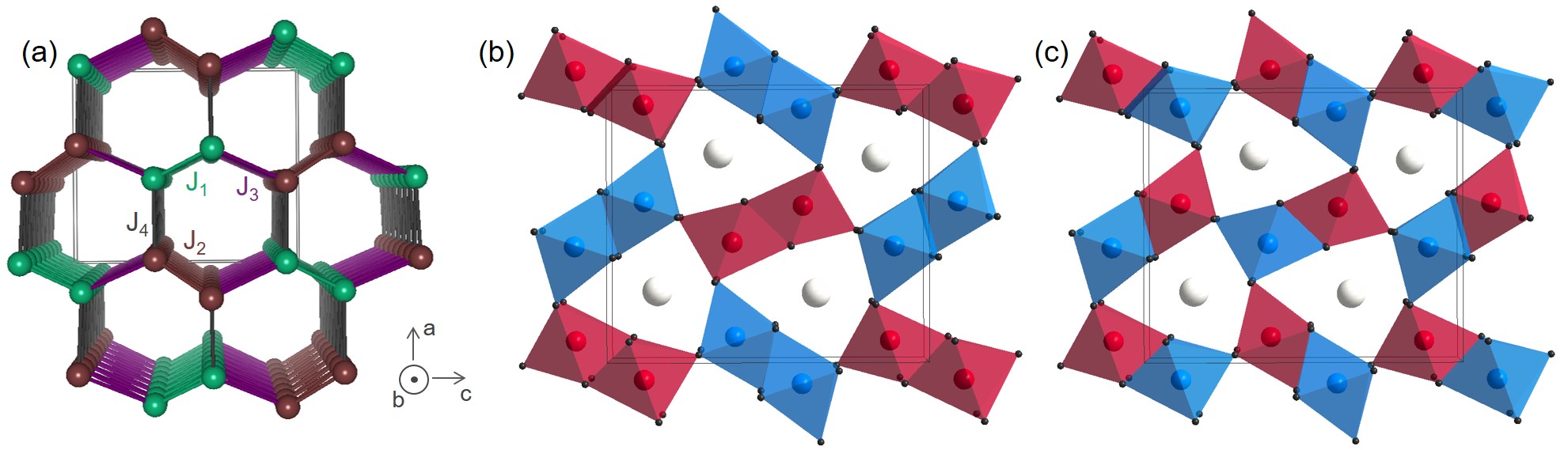}%
\caption{\label{fig:1}\textbf{Structure of CaFe\textsubscript{2}O\textsubscript{4} reproduced from the CIF file published by Galuskina \textit{et al.}\cite{Galuskina2014HarmuniteIsrael}} \textbf{a} Schematic representation of the distorted honeycomb lattice formed by $Fe$ atoms projected from the b-axis. The magnetic exchange is predominantly two dimensional with strong coupling ($J_3$ and $J_4$) along a and weak coupling ($J_1$ and $J_2$) along c. Green and brown colours indicate $Fe(1)$ and $Fe(2)$ sites, $Ca$ and $O$ atoms are here omitted. \textbf{b-c} Representation of the A and B spin structures with FM and AF intra-chain ($J_1$ and $J_2$) interactions, respectively. Blue and red colours indicate $Fe^{3+}$ spins parallel and antiparallel to the b-axis, $Ca$ atoms are represented in white and $O$ atoms in black.}
\end{centering}
\end{figure}

In the $CaFe_2O_4$ structure, the $Fe^{3+}$ ions occupy two crystallographically distinct positions, $Fe(1)$ and $Fe(2)$, each surrounded by 6 oxygen atoms in octahedral coordination, that form zig-zag chains that run parallel to the b-axis. $FeO_6$ octahedra within the same chain share edges, whereas neighbouring chains are connected through corners, as shown in Fig.\ref{fig:1} \cite{Das2018FerrimagneticCaFe2O4delta}. 
As in many oxides, the magnetic coupling between spins occurs via oxygen mediated superexchange, whose strength and sign depend on the $Fe-O-Fe$ bond angles. 
Thus, strong inter-chain antiferromagnetic (AF) interactions, $J_3$ and $J_4$, are found between corner sharing $Fe(1)O_6$ and $Fe(2)O_6$ octahedra, where the bond angles are about 120$^\circ$.
On the other hand, weaker couplings, $J_1$ and $J_2$, occur between edge-sharing $FeO_6$ octahedra within the same zig-zag chain, connected through angles of approximately 100$^\circ$. Recently, Das \textit{et al.}\cite{Das2018FerrimagneticCaFe2O4delta} have suggested that the magnetic structure of $CaFe_2O_4$ can be viewed as an armchair-type structure extending along the a-axis.

Below the Néel temperature two competing spin arrangements, named A and B, exist, that differ for the sign of the weak intra-chain couplings and, thus, on the c-axis stacking of $Fe^{3+}$ spins \cite{Watanabe1967NeutronCaFe2O4,Stock2016SolitaryCaFe2O4}. Specifically, the B structure is characterized by alternating spin up and spin down stripes in the c-direction, while in the A structure the periodicity is doubled with an up-up-down-down configuration (see Fig. \ref{fig:1}b-c). In both structures $Fe^{3+}$ spins align parallel to the b-axis, giving rise to an Ising-like system with large magnetocrystalline anisotropy \cite{Gandhi2017MagnetocrystallineCrystal}. At the Néel temperature ($T_N$=200 K) the material orders in a pure B phase. Upon decreasing temperature below 150 K, the A phase also appears and the coexistence of these two structures has been reported to occur down to low-temperatures, where the A arrangement is favoured \cite{Stock2017OrphanCaFe2O4, Corliss1967MagneticFerrite}. Interestingly, each phase can also be visualized as the local structure of the antiphase boundary between two domains of the other phase, where the ``orphan spins" generate an uncompensated magnetic moment along the b-axis\cite{Stock2017OrphanCaFe2O4}.\\

The magnetic properties of this material have been investigated by means of neutron diffraction, DC and AC magnetometry, on single-crystalline and polycrystalline samples. \cite{Apostolov1984InvestigationSystem,Lobanovsky2011MagneticSynthesis, Gandhi2017MagnetocrystallineCrystal,Stock2016SolitaryCaFe2O4, Das2018FerrimagneticCaFe2O4delta, Songvilay2020Disorderx=00.5}. However, there is no complete agreement in the literature on interpreting the magnetic susceptibility measurements. In particular, the magnetic properties of $CaFe_2O_4$ seem to be very sensitive to the oxygen content. For example, only one magnetic transition at lower $T_N$ has been observed in oxygen-deficient $CaFe_2O_4$ \cite{Das2018FerrimagneticCaFe2O4delta}. In addition, oxygen vacancies-driven partial conversion of $Fe^{3+}$ (HS S=5/2) into $Fe^{2+}$ (LS S=2) ions has been reported to cause incomplete cancellation of the magnetization below $T_N$ inducing ferrimagnetic behaviour. On the other hand, a ferrimagnetic state is also observed in oxygen superstoichiometric $CaFe_2O_4$ due to the presence of $Fe^{4+}$ ions and the charge disproportionation between $Fe^{3+}$ and $Fe^{4+}$ ions occupying two inequivalent sublattices\cite{Lobanovsky2011MagneticSynthesis}.\\

In this work we report a detailed structural and magnetic characterization of epitaxial thin films of $CaFe_2O_4$. The films are relaxed to the bulk structure and show magnetic properties consistent with those reported in single crystals \cite{Stock2016SolitaryCaFe2O4, Gandhi2017MagnetocrystallineCrystal}. The well defined microstructure of the films allows us to perform local magnetic characterization, yet unreported in this material, and to shed light into the origin of the net magnetic moment reported in various works \cite{Stock2017OrphanCaFe2O4, Das2018FerrimagneticCaFe2O4delta, Lobanovsky2011MagneticSynthesis}.
%%%%%%%%%%%%%%%%%%%%%%%%%%%%%%%%%%%%%%%%%%%%%%%%%%%%%%%%%%%%%%
\section*{Results}
%%%
\subsection*{Synthesis and crystal structure}
Finding a suitable substrate is the first step for the epitaxial growth of thin films. Unlike for Perovskite and Spinel-type materials, most of the commonly used crystalline substrates do not match the lattice parameters of the $CaFe_2O_4$ prototype structure, making predictions of the epitaxial relation between the $CaFe_2O_4$ film and substrate not straightforward. A previous work on thin films of this material has used $TiO_2$ (100) substrates \cite{Nishiyama2017HighlyDeposition}, due to the similarity between the oxygen octahedra in the rutile-type and $CaFe_2O_4$ structure. Thus, in our work, we also selected $TiO_2$ crystals as substrates, but cut along the (110) direction, in order to obtain a different out-of-plane orientation of the film.\\
The optimization of the growth of $CaFe_2O_4$ thin films on $TiO_2$ (110) substrates by Pulsed Laser Deposition (PLD) requires the control of several physical parameters (see Methods section). Because of the large nominal mismatch between film and this substrate orientation (9$\%$), polycrystalline or amorphous films are easily obtained for a large window of growth parameters. However, we observed that relatively thick films of around 100 nm, prepared with a number of laser pulses in between 6 and 20 thousand, as well as a high laser repetition rate (10-15 Hz), are crystalline and textured.\\

\begin{figure}[htb]
\begin{center}
\includegraphics[width=.8\linewidth]{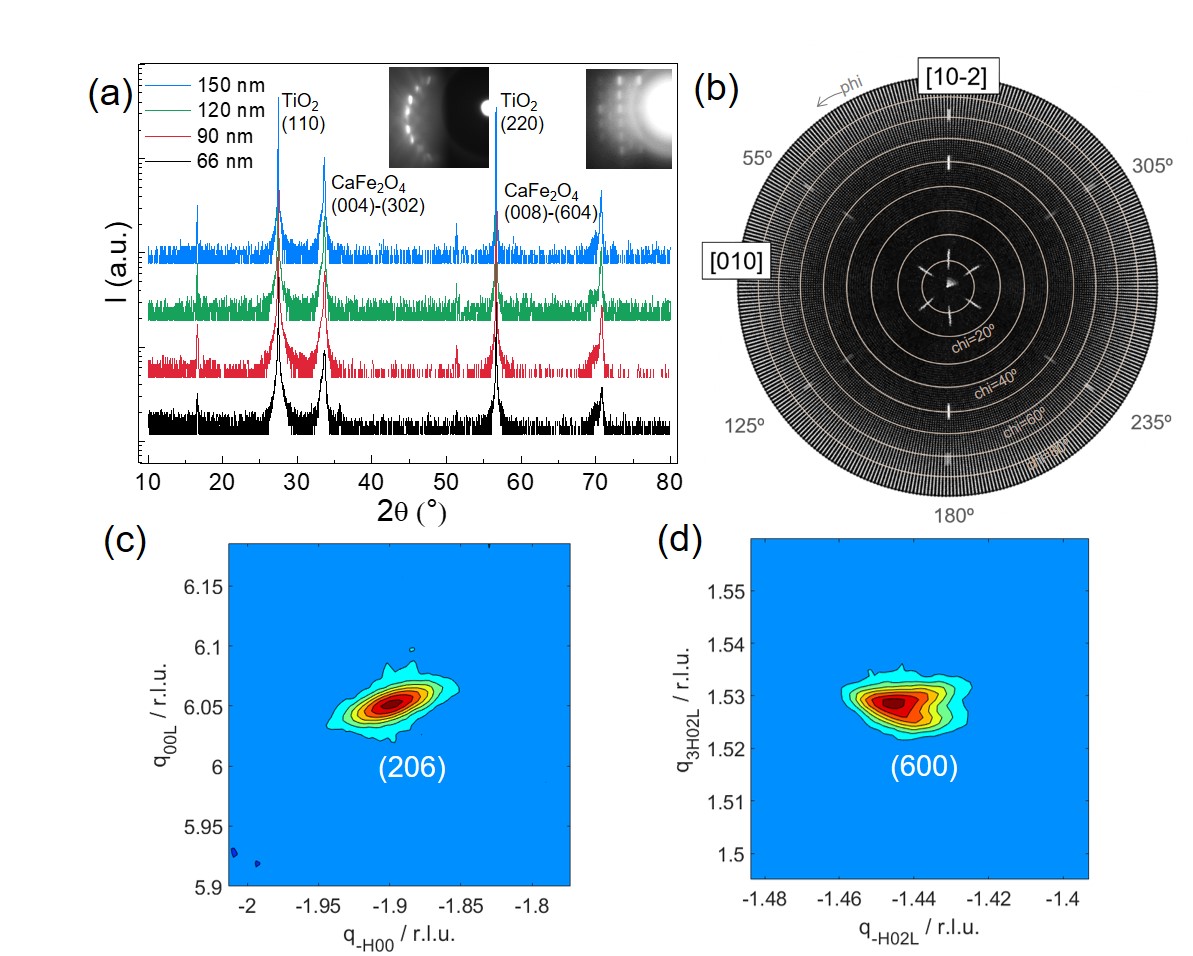}%
\caption{\label{fig:2}\textbf{Orientation determination via X-Ray Diffraction.} \textbf{a} Plot of the two-theta-omega scan from 10$^\circ$ to 80$^\circ$ for films of increasing thickness from 66 to 150 nm. In addition to the substrate peaks (2$\theta$=27$^\circ$ (110) and 2$\theta$=56$^\circ$ (220)) two film peaks are visible at 2$\theta$=33.6$^\circ$ and 2$\theta$=70.5$^\circ$. The insets show the RHEED patterns before and during the film deposition. \textbf{b} X-Ray pole figure taken at 2$\theta$=25.5$^\circ$ (202). The peaks at Chi=50$^\circ$ indicate the presence of 3 domains with (004) out-of-plane orientation, while those at $\chi$=10$^\circ$ and 70$^\circ$ originate from three (302) domains. \textbf{c-d} RSMs collected at $\chi$=$\phi$=0$^\circ$ showing the presence of the both the (-206) and (600) peaks, which is only possible if the (004) and (302) orientations coexist within the same film. Here, the r.l.u refer to bulk $CaFe_2O_4$ lattice constants.}
\end{center}
\end{figure}

Following the films growth \textit{in-situ} by reflection high energy electron diffraction (RHEED) indicates island-growth mode: during the first minutes of deposition, the initial sharp reciprocal rods of the atomically flat substrate evolve into a transmission diffraction pattern typical of 3D islands \cite{Hasegawa2012ReflectionDiffraction}. Finally, at the end of the deposition, no more rods are visible, indicating high surface roughness (see inset Fig. \ref{fig:2}a). Despite this, a well-defined epitaxial relation between the films and the substrate is observed, as discussed below.

Increased crystallinity of the films, estimated by the intensity of the out-of-plane peak in the X-Ray diffraction (XRD) pattern (Fig. \ref{fig:2}a), was achieved with a substrate temperature of 850 \si{\celsius} and partial oxygen pressure $P_{O2}$=0.2 mbar. A relatively high energy density of 2.8 J/cm\textsuperscript{2} was required in order to ablate $Fe$ and $Ca$ atoms in equal proportion from the ceramic target and achieve near stoichiometric transfer (see Supplementary Note 1). As a result, $Ca$ atoms travelling in the plasma plume reach the $TiO_2$ surface with high energy and are able to interact chemically with it. This leads to the formation of a Calcium Titanate layer at the interface between film and substrate.
\\

Fig. \ref{fig:2} shows the characterization of the films by means of X-Ray Diffraction (XRD). Two strong peaks in the two-theta-omega scans (Fig. \ref{fig:2}a) are seen at angles of 33.6$^\circ$ and 70.5$^\circ$. The former can belong to both the (004) and (302) planes of $CaFe_2O_4$ and the latter to their second order diffraction. These two families of planes not only share the same lattice spacing, $d=2.67$ \AA, but also display a very similar arrangement of atoms, making it non-trivial to tell them apart in X-Ray experiments (for more details see Supplementary Note 2). Therefore, in order to precisely determine the films orientation, the data from specular reflections need to be complemented by Reciprocal Space Maps (RSMs) around off-specular peaks. In the first map (Fig. \ref{fig:2}c), we observe a peak at 2$\theta$=55.23$^\circ$ and $\omega$=6.48$^\circ$, which is the (-206) peak if (004) is the out-of-plane orientation. No peak should be observed in that position in case of the (302) orientation. In the second map (Fig. \ref{fig:2}d), we observe a peak at 2$\theta$=60.25$^\circ$ and $\omega$=-1$^\circ$, which is the (600), if (302) is the out-of-plane orientation. Again, no peak should be found at these position in case of the (004) orientation. Therefore, the presence of both the (-206) and (600) peaks is only consistent with the coexistence of both (004) and (302) out-of-plane orientations within the same film. Moreover, from the RSMs we can deduce the epitaxial relation between films and substrate. In both crystal orientations, the [010] direction of the film is in-plane and aligned with the [1-10] direction of the substrate. On the other hand, the substrate [001] direction is parallel to the [100] and [10-2] directions of $CaFe_2O_4$ for (004) and (302) oriented crystals, respectively. This is particularly relevant for the magnetic properties of the films, being the [010] (b-axis) the magnetization direction, which indicates that the $Fe^{3+}$ spins are oriented in the plane of the films.

Further proof of the coexistence of the (004) and (302) orientations is provided by X-Ray pole Figures. Fig. \ref{fig:2}b shows the measurement collected at 2$\theta$=25.5$^\circ$ that corresponds to the lattice spacing of the (202) planes of $CaFe_2O_4$. The normal to such planes forms an angle of 50$^\circ$ with the [004] direction and 10$^\circ$ with the [302] direction. Therefore, 2 peaks (at $\phi$=90$^\circ$ and 270$^\circ$ from the [010] direction) are expected to appear when rotating the sample with respect to the film normal, for $\chi$= 50$^\circ$ and 10$^\circ$. In Fig. \ref{fig:2}d, six peaks for each value of $\chi$ appear, indicating that both orientations exist and each of them contains three domains (see the next section). Moreover, we also observe six peaks at $\chi$=70$^\circ$,  corresponding to the (-103) planes, with a d-spacing close to that of the (202) planes, forming a 70$^\circ$ angle with the (302) planes.\\

\begin{figure}[htb]
\begin{center}
\includegraphics[width=.98\linewidth]{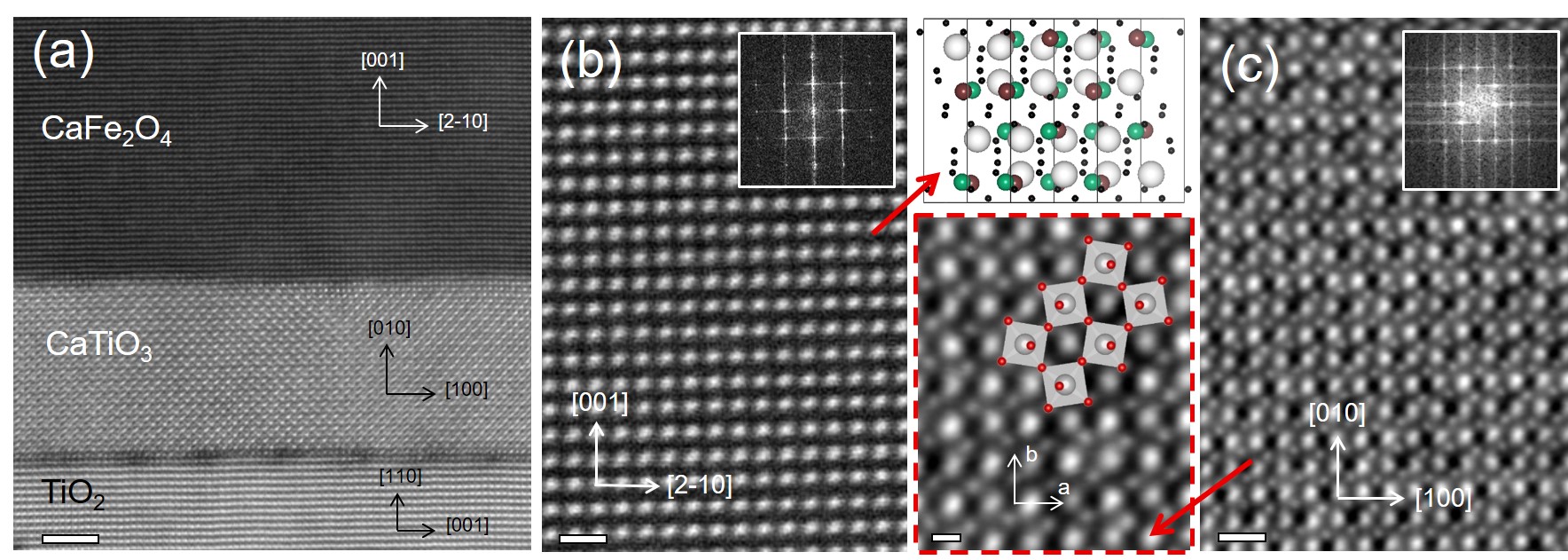}%
\caption{\label{fig:3}\textbf{Transmission electron microscopy (TEM).} Cross-section images of a 90 nm thick $CaFe_2O_4$ film with an intermediate $CaTiO_3$ layer of about 7 nm imaged along the [120] and [001] zone axes, respectively. \textbf{a} HAADF-STEM overview image showing the $TiO_2$ substrate surface, $CaTiO_3$ layer and $CaFe_2O_4$ film on top. The scale bar has length 2.5 nm. \textbf{b} HAADF-STEM magnification of the $CaFe_2O_4$ lattice. In the inset the FFT, from which the out-of-plane lattice parameter is measured, is shown. The second inset depicts a model of the crystal structure imaged with the same orientation, evidencing the square-like pattern formed by columns of $Fe(1)$ (green), $Fe(2)$ (brown) and $Ca$ (white) atoms. \textbf{c} iDPC-STEM magnification of the Perovskite $CaTiO_3$ lattice. In the inset the FFT is shown, from which the in-plane and out-of-plane lattice parameters are measured. The second inset shows the $TiO_6$ octahedra tilt imaged along the $CaTiO_3$ [001] direction, revealing the a-a-c+ oxygen octahedral tilt pattern characteristic of the $Pnma$ space group; $Ca$ atoms appear in white and $O$ in red. The scale bars in \textbf{b-c} have length 500 pm and in the inset 200 pm.}
\end{center}
\end{figure}

The local structure of the films was further analysed by Transmission Electron Microscopy (TEM) (Fig. \ref{fig:3}). High angle annular dark field scanning TEM (HAADF-STEM) and corresponding energy dispersive spectroscopy (EDS) analysis revealed the presence of a 10 nm $CaTiO_3$ layer with the perovskite structure between the substrate and the $CaFe_2O_4$ film, arising out of a chemical reaction between the high energy Ca$^{2+}$ ions in the plasma and the $TiO_2$ substrate surface (see Supplementary Note 1). The $CaTiO_3$ layer is (010) oriented, and fully relaxed by means of dislocations, with 6 planes of the films corresponding to 5 planes of $TiO_2$ (inset of Fig. \ref{fig:3}c).

HAADF-STEM image of the $CaFe_2O_4$ layer is shown in Fig. \ref{fig:3}b. The square-like pattern corresponds to the projection from the [120] zone axis of a crystallite with (004) out-of-plane orientation. The in-plane lattice parameter of d=2.53\si{\angstrom}$^{-1}$ corresponds to the (210) d-spacing. This indicates that, in the crystal imaged here, the $CaFe_2O_4$ [010] direction is tilted with respect to the to substrate [1-10] by an angle of approximately 55$^\circ$. This is consistent with the domain structure observed by means of Atomic Force Microscopy (AFM) and discussed in the next section.

The oxygen column imaging was further performed through differential phase contrast (DPC) STEM. The integrated DPC-STEM image on the $CaTiO_3$ layer (Fig. \ref{fig:3}c and corresponding inset) clearly reveals a-a-c+ oxygen octahedral tilt pattern, corresponding to orthorhombic Pnma symmetry. Furthermore, the $CaTiO_3$ layer also exhibits 178$^\circ$ ferroelastic domain boundaries, reminiscent of bulk $CaTiO_3$\cite{VanAert2012DirectMicroscopy} (for more details see Supplementary Note 3). 
%%%%%%%%%%%%%%%%%%%%%%%%%%%%%%%%%%%%%%%%%%%%%%%%%%%%%%%%
\subsection*{Domain Structure}
\begin{figure}[htb]
\begin{centering}
\includegraphics[width=.7\linewidth]{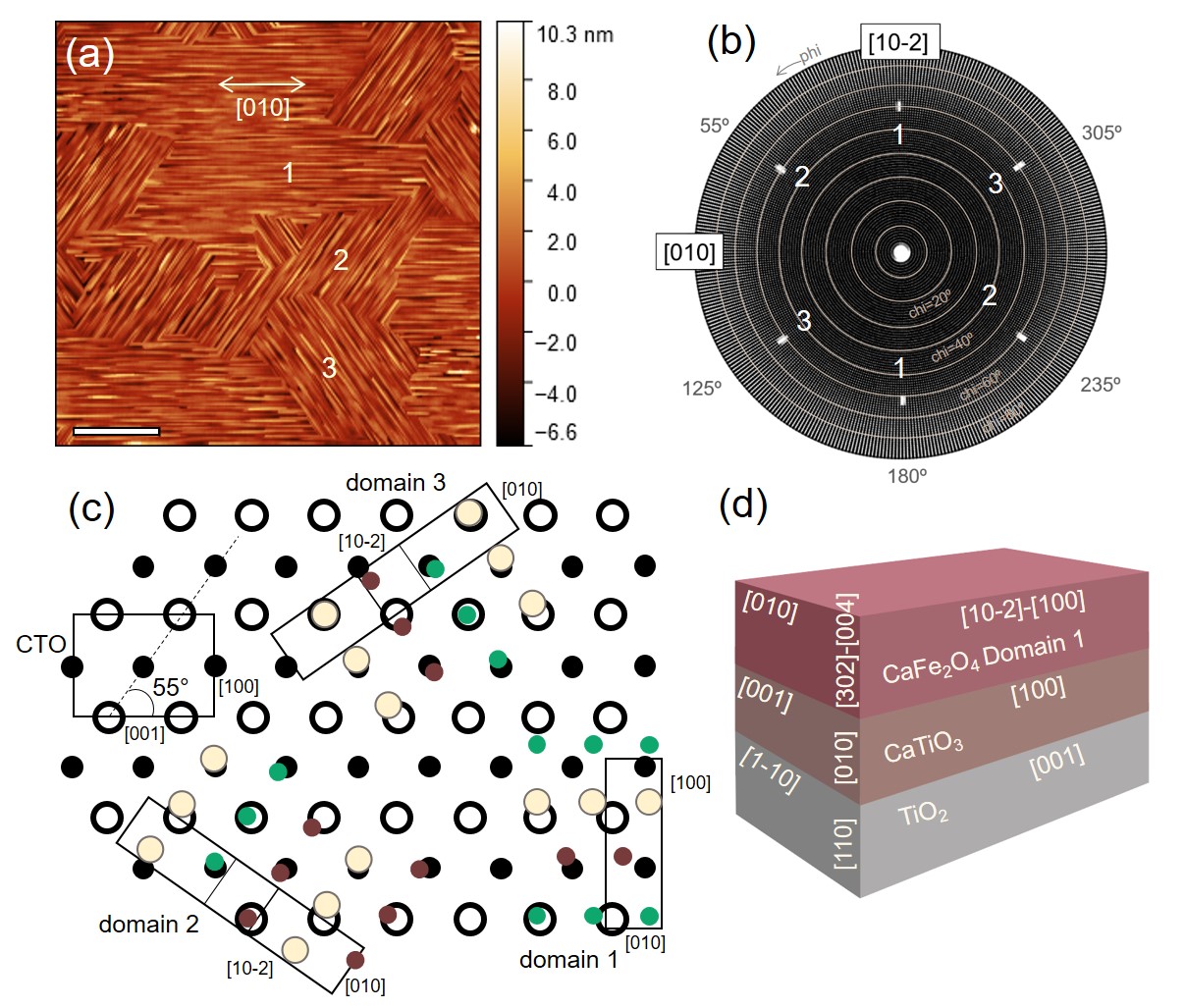}%
\caption{\label{fig:4}\textbf{Domain structure.} \textbf{a} AFM image (topography) of a 90 nm thick sample. 1,2 and 3 indicate the three possible domain orientations in the film. The scale bar has length 3 $\mu$. \textbf{b} Pole figure collected at 2$\theta$=33.6$^\circ$ with the substrate [1-10] parallel to the scattering plane. Here six spots are visible at $\chi$=60$^\circ$ and $\phi$=0$^\circ$, 55$^\circ$, 125$^\circ$, 180$^\circ$, 235$^\circ$ and 305$^\circ$. \textbf{c} Schematic representation of the epitaxial relation between different domains of the $CaFe_2O4$ film and under-laying $CaTiO_3$ layer. $Fe(1)$ is shown in green, $Fe(2)$ in brown and $Ca$ in white. The black circles indicate the cations in the underlying $CaTiO_3$ layer, with empty circles for $Ca$ positions and filled circles for $Ti$. \textbf{d} Cartoon displaying the proposed epitaxial relationship between domain 1 of $CaFe_2O4$ film, $CaTiO_3$ layer and $TiO_2$ substrate.}
\end{centering}
\end{figure}

The $CaFe_2O_4$ thin films prepared in this study display a distinctive domain structure, as clearly seen in the images collected by means of Atomic Force Microscopy (AFM). Each domain is composed of needle-like crystallites with the long axis parallel to the [010] direction. Three specific crystallographic orientations of the domains are found as shown in Fig. \ref{fig:4}a: 1- with the [010] parallel to the substrate [1-10], 2 - forming a 55$^\circ$ angle with 1 and 3 -forming a -55$^\circ$ angle with 1. Consistent results are obtained from X-Ray pole figure measurements. Fig. \ref{fig:4}b shows the data collected at 2$\theta$=33.65$^\circ$, that corresponds to the spacing of $CaFe_2O_4$ (302) and (004) planes (first film peak in the 2theta-omega scan of Fig. \ref{fig:2}a). Here, for a single domain sample, two peaks are expected to appear at $\chi$=60$^\circ$ and $\phi$=90$^\circ$, 270$^\circ$ from the [010] direction. However, together with these, we observe 4 more peaks at $\phi$=55$^\circ$, 125$^\circ$, 235$^\circ$ and 305$^\circ$, which indicate the presence of 3 $CaFe_2O_4$ domains. Finally, the same domain structure emerges when studying the films by means of Electron Backscattered Diffraction (EBSD) in a scanning electron microscope (SEM), which allows to determine the crystallites orientation (see Supplementary Note 4).\\
To explain the formation of 55$^\circ$ domains in the above mentioned directions, we put forward a model based on optimum structural matching between the crystal lattice of $CaFe_2O_4$ and that of the underlying $CaTiO_3$ layer. We notice that 55$^\circ$ is the angle between the $CaTiO_3$ [001] and [101] in-plane directions. The arrangement of the atoms in the (302) and (004) planes of $CaFe_2O_4$ consists of similarly spaced rows of cations that run parallel to the [010] direction. In both cases, two $Fe$ rows alternate with one $Ca$ row. As Fig. \ref{fig:4}c shows, the atoms belonging to the two layers overlap best when the cations rows of $CaFe_2O_4$ are either parallel to the $CaTiO_3$ [001] direction or at $\pm$55$^\circ$ from it. Because the growth of the films of this study follows an island-growth mode, islands with one of the 3 orientations start growing independently and later merge together yielding a rough film. The boundary between 2 adjacent domains is sharp with an herringbone pattern, whereas at the conjunction between 3 or more crystallites, vortex-like structures that can have triangular or diamond shape, are visible. A cartoon to better illustrate the complex epitaxial relation between film and substrate, comprehensive of $CaTiO_3$ intermediate layer, is shown in Fig. \ref{fig:4}d.

%%%%%%%%%%%%%%%%%%%%%%%%%%%%%%%%%%%%%%%%%%%%%%%%%%%%%%%%%%%%
\subsection*{Magnetic properties}
\begin{figure}[htb]
\begin{centering}
\includegraphics[width=\linewidth]{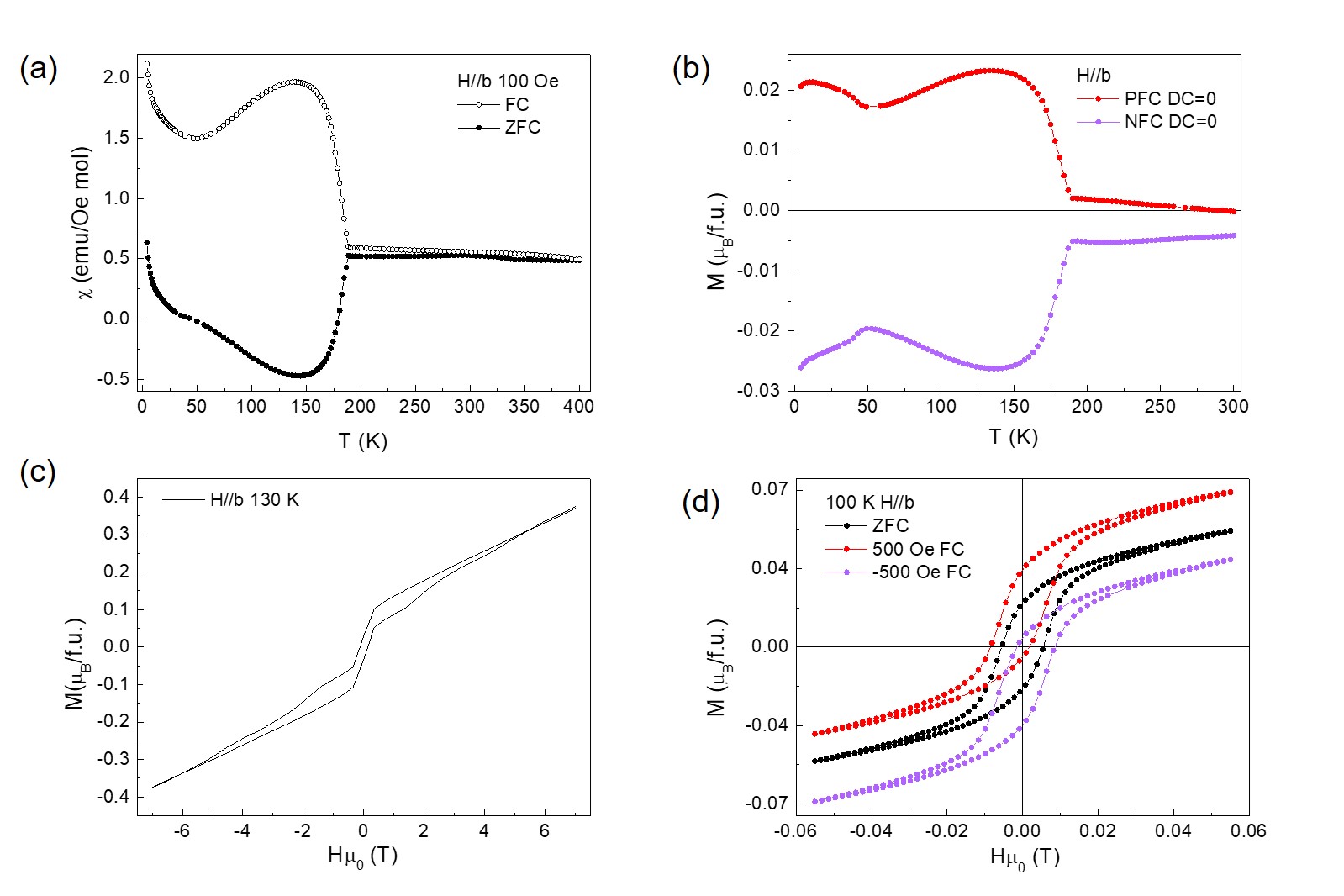}%
\caption{\label{fig:5} \textbf{Magnetic properties of CaFe\textsubscript{2}O\textsubscript{4} thin films measured with field parallel to the magnetization direction (b-axis).} \textbf{a} Plot of the magnetic susceptibility ($\chi$) of a 84 nm thick sample as a function of temperature ($T$) from 5 to 400 K in a 100 Oe magnetic field. \textbf{b} Magnetization ($M$) of a 90 nm thick sample measured as a function of $T$ from 5 to 300 K in zero applied DC field after field cooling under 100 Oe (red) and -100 Oe (purple).\textbf{c} Magnetization ($M$) of a 96 nm thick sample as a function of applied field ($H$) measured at 130 K between 7 and -7 T. \textbf{d} Magnetization ($M$) of a 90 nm thick sample as a function of applied field ($H$) measured at 100 K between 550 and -550 Oe after ZFC (black), 500 Oe FC (red) and -500 Oe FC (purple).}
\end{centering}
\end{figure}

After optimization of the growth process, we investigated the magnetic properties of $CaFe_2O_4$ thin films at both local and macro scales. The magnetization of the films is measured as a function of temperature using a SQUID magnetometer for different values of applied magnetic field ($H$). The magnetic susceptibility ($\chi=M/H$) from 4 to 400 K in a 100 Oe field parallel to the magnetization direction (b-axis of $CaFe_2O_4$) is plotted in Fig. \ref{fig:5}a. Here, a clear transition is observed at $T_N$=188 K (determined by the onset of DC magnetization), where $\chi$ steeply increases in the field-cooled (FC) curve and decreases in the zero-field-cooled (ZFC) one.
Upon decreasing temperature, $\chi$ reaches a maximum at $T$=140 K after FC, while at the same temperature, $\chi$ reaches a minimum after ZFC. The noticeable splitting of the FC and ZFC data, also observed in our ceramic PLD target (see Supplementary Note 5), evidences the presence of a ferrimagnetic contribution added to the expected AF response. Moreover, in the films case, a small ZFC/FC splitting persists up to temperatures above $T_N$, where the magnetization value is non-zero. This could be due to the remanent fields that are unavoidably present in the SQUID magnetometer, with different sign depending on the history of the previously applied field\cite{Buchner2018Tutorial:Spintronics, Kumar2010OnProcess}. 

In addition, differently from bulk, in the $\chi$ \textit{vs.} $T$ plots (Fig. \ref{fig:5}a) a paramagnetic (PM) tail can be found at below 30 K, that can probably be attributed to the $CaTiO_3$ layer at the interface between films and substrates (the latter being diamagnetic). Moreover, the magnetic susceptibility of $CaFe_2O_4$ thin films shows strong orientation dependence, being noticeably lower when the applied magnetic field is perpendicular to the b-axis (see Supplementary Figure 6a-b). This indicates strong magnetocrystalline anisotropy, which is expected for an Ising-like system as $CaFe_2O_4$ \cite{Gandhi2017MagnetocrystallineCrystal}.

To further investigate the ferrimagnetic behaviour of $CaFe_2O_4$, we measured the magnetization ($M$) as a function of temperature ($T$) in zero applied field. Fig. \ref{fig:5}b shows the data collected after cooling in a field of $\pm$100 Oe parallel to the b-axis. The measured magnetic response indicates the presence of a spontaneous magnetization in $CaFe_2O_4$ films. On the other hand, here the low temperature tail observed in Fig. \ref{fig:5}a is absent, confirming its paramagnetic nature. Next to the ordering temperature at $T_N$=188 K, an anomaly at around 35 K and a broader feature above 200 K are also visible. Such features were also observed in previous studies and have been assigned to a slow spin dynamical process \cite{Das2018FerrimagneticCaFe2O4delta} and room-temperature spin interactions \cite{Gandhi2017MagnetocrystallineCrystal, Das2018FerrimagneticCaFe2O4delta}, respectively.

The presence of an uncompensated magnetic moment is also supported by the hysteresis of the $M-H$ loops measured at various temperatures. In Fig. \ref{fig:5}c the measurement at 130 K is shown, where the maximum hysteresis is observed (see Supplementary Figure 6c for the data at 30 and 175 K). Furthermore, when the sample is cooled down through $T_N$ in the presence of a magnetic field parallel to the b-axis, the loop is subjected to a vertical shift in the direction of the applied field. Such shift is absent if the field is applied perpendicular to the magnetization direction.\\
Measuring $M-H$ loops at low fields (up to 500 Oe) also reveals a small hysteresis that persists above $T_N$, but no induced shift is observed under FC conditions (see Supplementary Figure 6d).\\

\begin{figure}[htb]
\begin{centering}
\includegraphics[width=.72\linewidth]{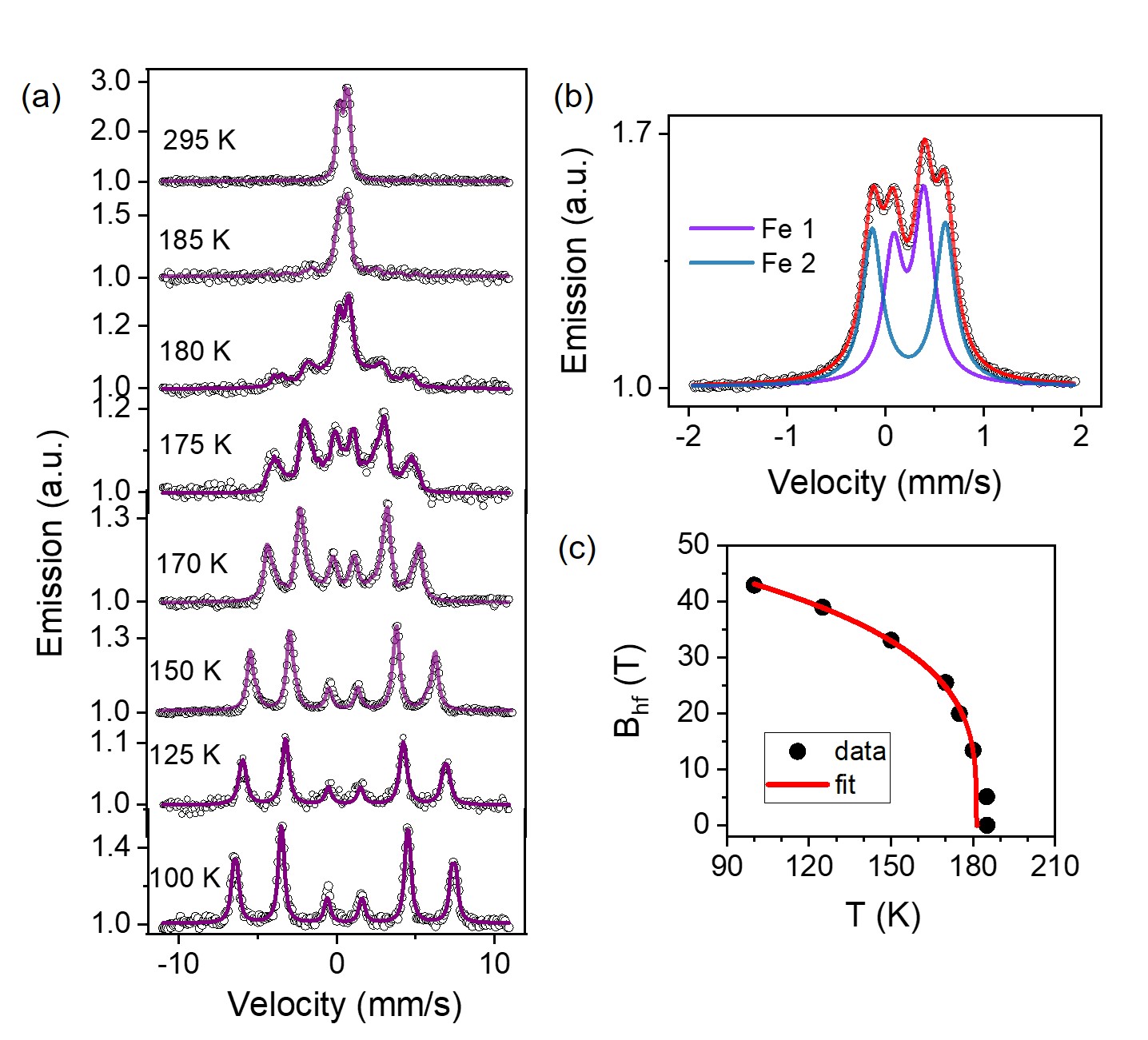}%
\caption{\label{fig:6} \textbf{Mössbauer spectra of \textsuperscript{57}Fe-enriched CaFe\textsubscript{2}O\textsubscript{4} thin films.} \textbf{a} Conversion electron Mössbauer spectra at temperatures ranging between 300 K and 100 K. \textbf{b} CEMS spectrum at 300 K recorded in a narrow velocity range. \textbf{c} Temperature dependence of the hyperfine field. The solid line corresponds to the fit with a power law behavior.}
\end{centering}
\end{figure}

In order to further characterize the magnetic structure of $CaFe_2O_4$ films, investigate the oxidation state of $Fe$ and rule out the possibility of contamination with different $Fe$-containing phases or oxides, we also performed Mössbauer Spectrometry in electron conversion mode (CEMS) (Fig. \ref{fig:6}). The room-temperature CEMS spectrum (Fig. \ref{fig:6}a) exhibits a sharp paramagnetic doublet without any trace of magnetic parasitic phases containing $Fe$. Therefore, we can exclude contamination by iron oxides or other calcium ferrite phases with higher $T_N$, such as brownmillerite $Ca_2Fe_2O_5$\cite{Kagomiya2017WeakGradient} or $CaFe_3O_5$\cite{Hong2018LongCaFe3O5, Cassidy2019SingleOrdering}. A high resolution CEMS spectrum recorded at RT in a narrow velocity scale is reported in Fig. \ref{fig:6}b. This spectrum shows well-defined lines and was fitted with two paramagnetic quadrupolar doublets corresponding to the two inequivalent $Fe^{3+}$ sites $Fe(1)$ and $Fe(2)$, as expected for a pure $CaFe_2O_4$ phase\cite{Hudson1967MossbauerPoint, Yamamoto1968MossbauerCaFe2O4, Hirabayashi2006MossbauerCaO,Tsipis2007Oxygen/sub,Kharton2008MixedCaFe2O4-delta, Berchmans2010AStructure}. Both components have almost equal spectral area and linewidths (full width at half maximum $\Gamma\backsim$0.24 mm s\textsuperscript{-1}). The isomer shift values are also similar ($\delta$=0.368$\pm$0.001 mm s\textsuperscript{-1}), but the quadrupole splitting ($\Delta$ $E_Q$) is different, with values of 0.313$\pm$0.001 mm s\textsuperscript{-1} and 0.743$\pm$0.001 mm s\textsuperscript{-1} for $Fe(1)$ and $Fe(2)$, respectively. 

The isomer shift values are typical of $Fe^{3+}$ ions, and the absence of signal belonging to $Fe^{2+}$ suggests low oxygen vacancy content in the film. An asymmetry of the line intensity of the doublet, different for each site, is clearly evidenced. Such asymmetry, in case of single crystal and isotropic Lamb-Mössbauer factor, is due to a preferred orientation of the symmetry axis of the electric field gradient (EFG) at the nucleus. If the principal axis of the EFG makes an angle $\theta$ with the incident $\gamma$-beam direction, the line intensity ratio of the quadrupolar doublet is given by $I_2/I_1 =3(1+\cos^2\theta)/(5-3\cos^2\theta)$, with values ranging from 3 for $\theta$=0 to 0.6 for $\theta$=90$^\circ$. Here the fit of the spectrum yields $\theta$= 41$^\circ$ and 53$^\circ$ for $Fe(1)$ and $Fe(2)$, respectively. 

In Fig. \ref{fig:6}a also some selected CEMS spectra at temperatures below room-temperature are reported. The CEMS spectra below 185 K clearly show the onset of long range magnetic order by the appearance of a magnetic sextet due to nuclear Zeeman splitting. For each temperature, the line intensity ratios are close to 3:4:1:1:4:3 for the magnetic sextet, evidencing in plane orientation of the $Fe$ spins. The temperature dependence of the mean magnetic hyperfine field $B_{hf}$ deduced from the fit can be approximated using a power law $B_{hf}(T) = B_{hf}(0) (1-T/T_N)^{\beta}$, where $\beta$ is the critical exponent or the AF order parameter (the staggered sub-unit cell magnetization). A reasonably good fit (Fig. \ref{fig:6}c) leads to $B_{hf}(0)$= (54.8$\pm$4.0) T, $\beta$= 0.28$\pm$0.05, and $T_N$= (181.2$\pm$1.6) K. The value of the critical exponent is consistent with the $\beta$= 1/3 value expected for a 3D Ising antiferromagnet. The Néel temperature obtained from the fit is also consistent with the transition temperature deduced from the SQUID measurements.\\

\begin{figure}[htb]
\begin{centering}
\includegraphics[width=.65\linewidth]{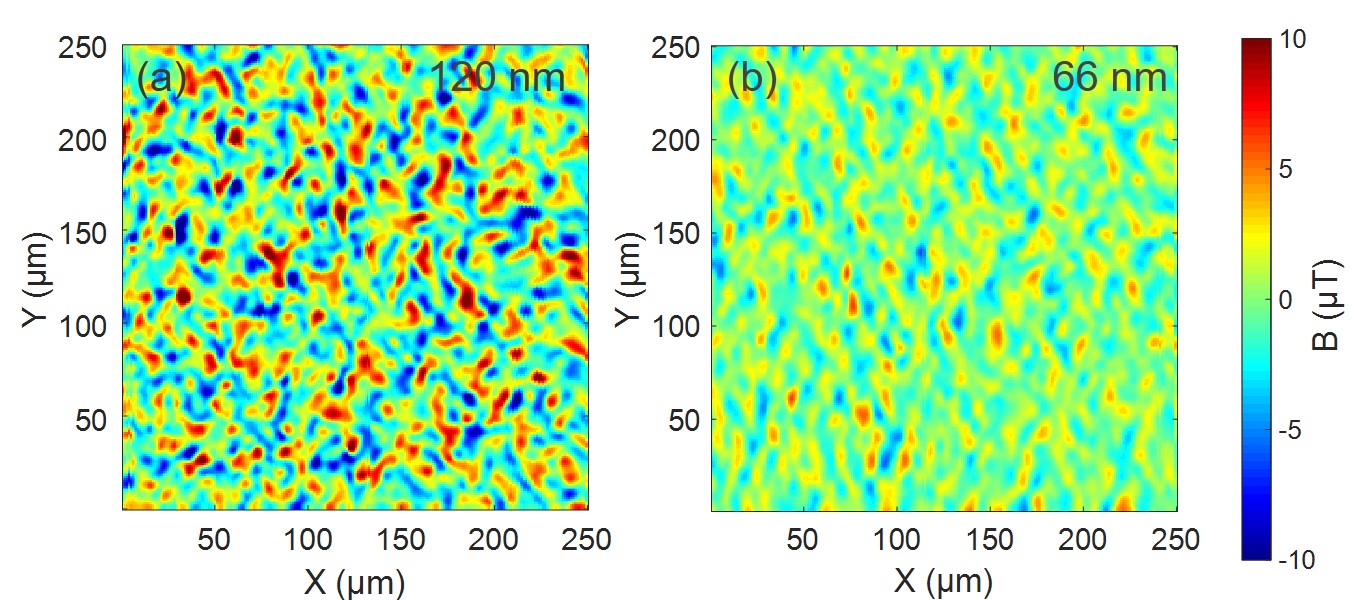}%
\caption{\label{fig:7} \textbf{Local magnetic response: scanning SQUID microscopy.} 250 $\mu$m x 250 $\mu$m scan of a \textbf{a} 120 nm film and \textbf{b} 66 nm film measured at 4 K.}
\end{centering}
\end{figure}

The local magnetic response of the $CaFe_2O_4$ films was also studied by means of scanning SQUID microscopy. Scans collected at 4 K (Fig. \ref{fig:7}) indicate clear magnetic activity. The observed patterns resemble those of a weak ferromagnet \cite{Reith2017AnalysingMicroscopy}, but no clear structure in the signal is visible. This is due to the spatial resolution of the scanning SQUID setup (approximately 5 $\mu$m) that causes averaging over multiple domains.
Different sample thicknesses give rise to similar magnetic patterns but with different intensities: for a 120 nm film (Fig. \ref{fig:7}a) the magnetic field measured is 7-8 $\mu$T, while when the thickness is reduced to 66 nm the field is approximately halved (Fig. \ref{fig:7}b). These values are well above the scanning SQUID sensitivity of approximately 50 nT. This confirms that the signal originates from the full $CaFe_2O_4$ film, and is not just limited to the surface.

\begin{figure}[htb]
\begin{centering}
\includegraphics[width=.75\linewidth]{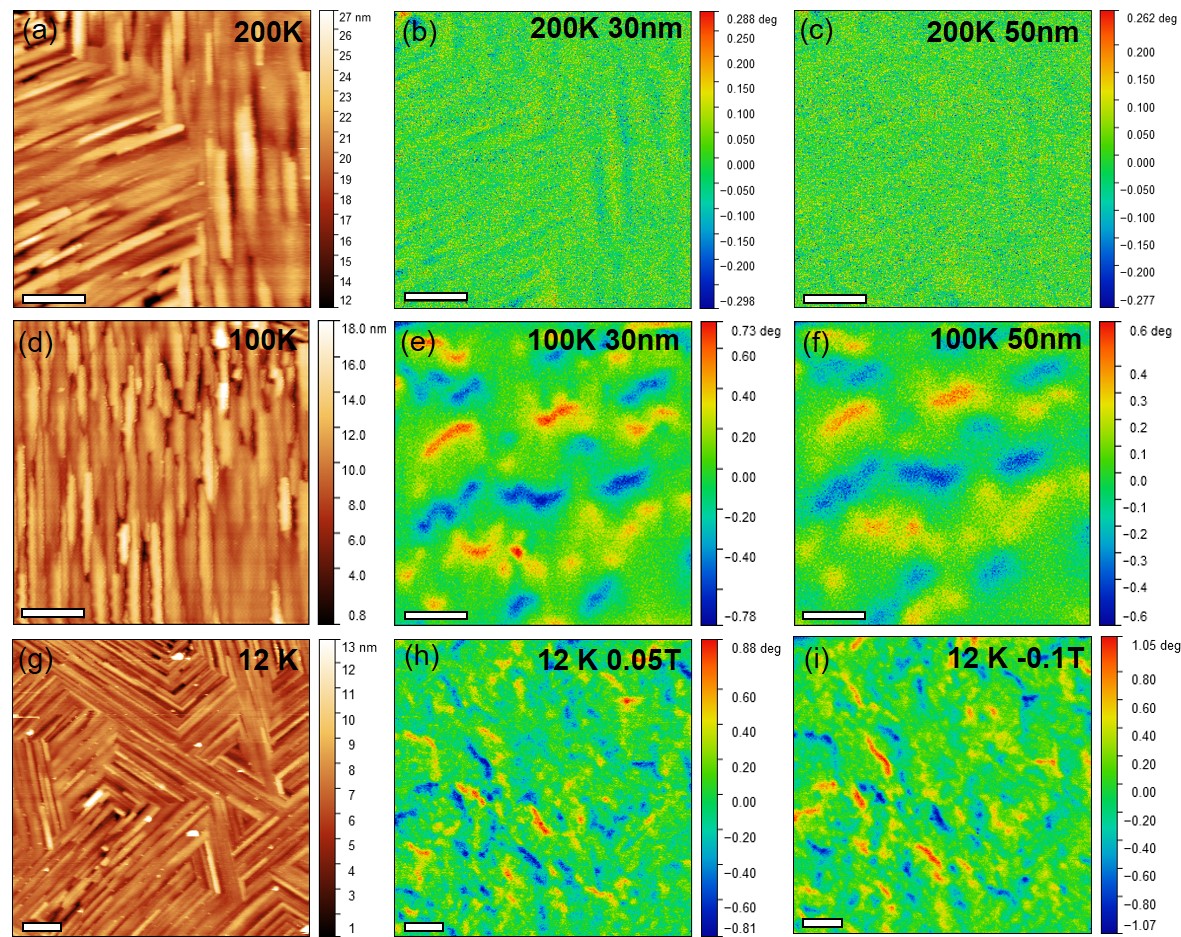}%
\caption{\label{fig:8}\textbf{Local magnetic response: low temperature MFM.} Images of a 120 nm thick sample. \textbf{a} Topography \textbf{b} dual-pass phase at 30 nm and \textbf{c} 50 nm lift measured at 200 K, above $T_N$. \textbf{d} Topography, \textbf{e} dual-pass phase at 30 nm lift and \textbf{f} at 50 nm lift measured at 100 K, below $T_N$. \textbf{g} Topography, \textbf{h} dual-pass phase at 30 nm lift with 0.05 T applied field and \textbf{i} dual-pass phase at 30 nm lift with -0.1 T applied field. All the scale bars have length 2 $\mu$m.}
\end{centering}
\end{figure}

In addition, in order to directly compare the magnetic and topographic features of the samples, we also performed magnetic force microscopy (MFM) experiments, that yields a spatial resolution of about 100 nm (Fig. \ref{fig:8}). Topography and MFM phase were recorded at various temperatures between 300 and 12 K, with a lift of either 30 nm and 50 nm from the sample surface.\\ The first images, collected from room-temperature down to 200 K (see Fig. \ref{fig:8}a-b-c) do not show any magnetic response. Here, the low contrast observed in Fig. \ref{fig:8}b can be attributed to simple cross-talk with the film topography, as an analogous signal is observed when the experiment is repeated with a non-magnetic tip, as shown in Supplementary Figure 7a-b.\\
Only when the temperature is lowered below the material's $T_N$ of 185 K a sharp contrast in the phase signal appears. Fig. \ref{fig:8}d-e-f show scans collected at 100 K. In these images we observe signatures of magnetic dipoles (alternating red and blue contrast), several of which seem to correspond to some of the edges of the needle-like crystals. Such signal increases in intensity and sharpness at lower scan lifts.
Fig. \ref{fig:8}g-h-i also show MFM images collected at 12 K in an applied magnetic field. Here, the color contrast in the second-pass phase is inverted upon reversing the magnetic field sign, from 0.05 T in \ref{fig:8}h to -0.1 T in \ref{fig:8}i (the difference between the two images can be seen in Supplementary Figure 7c). This indicates that the interaction between the tip and the sample goes from ferromagnetic to antiferromagnetic, and vice versa, upon reversing the tip magnetization.\\
These results are in good agreement with the expected scenario, in which the $Fe^{3+}$ spins align along the [010] direction that lies in the plane of the films. Such direction corresponds to the long axis of the needle-like domains, thus the magnetic field lines are only picked-up in MFM experiments (with sensitivity limited to out-of-plane magnetization) at the end of the crystallites, where the magnetic field lines bend in the out-of-plane direction. These results are also consistent with the SQUID measurements, showing that $CaFe_2O_4$ thin films do not display the pure AF behaviour.

%%%%%%%%%%%%%%%%%%%%%%%%%%%%%%%%%%%%%%%%%%%%%%%%%%%%%%%%%%%%%%%%%&&
\section*{Discussion}
Despite the single out-of-plane peak observed by XRD in the two-theta-omega scans, in-depth characterization reveals the coexistence of two crystal orientations with identical lattice spacing, namely (004) and (302). Distinguishing between such orientations is complicated by the similar arrangement of $Ca$ and $Fe$ atoms in these two families of crystal planes. The similarity between these two orientations combined with the high frequency deposition, causes islands of both to nucleate at the surface and merge in an homogeneous film as thickness increases. TEM characterization also reveals that the epitaxial growth of $CaFe_2O_4$ films is achieved through the formation of a perovskite $CaTiO_3$ layer at the interface with the $TiO_2$ substrate. The presence of this layer explains the domain structure of the films: oriented needle-like crystallites connected together by herringbone walls. We explain this in terms of optimum matching between the cation positions in the $CaFe_2O_4$ and $CaTiO_3$ lattices, which is achieved when the film [010] direction is parallel to the $CaTiO_3$ [001] (which is in turn epitaxial with the substrate [1-10]) or at $\pm$ 55$^\circ$ from it (see Fig. \ref{fig:4}). The presence of these domain variants gives rise to vortex-like structures. Interestingly, the magnetic easy axes of the two crystal orientations coincide, as well as the direction of the net magnetic moment at the antiphase boundaries \cite{Stock2017OrphanCaFe2O4}.\\

As expected for an Ising-like system, the magnetic response of $CaFe_2O_4$ films studied by means of SQUID magnetometry, displays a strong orientation dependence, being higher when the magnetic field is parallel to the b-axis of the crystals (comparison between Fig. \ref{fig:5}a-b and Supplementary Figure 6a-b). The behaviour of the magnetic susceptibility as a function of temperature (Fig. \ref{fig:5}a) is characterized by a single magnetic transition, defined as the onset of DC magnetization, which occurs at $T_N$= 188 K, and a maximum around T=140 K. In addition, fitting the hyperfine field thermal variation from the Mössbauer data gives rise to a $T_N$ $\approx$ 185 K as the only detectable transition. A single ground state (A phase) for the undoped material is in agreement with the phase diagram by Corliss \textit{et al.}\cite{Corliss1967MagneticFerrite} and a recent report by Songvilay \textit{et al.}\cite{Songvilay2020Disorderx=00.5}.

Another distinctive feature of the $\chi$ vs $T$ plots is the splitting of the FC and ZFC curves below $T_N$, with the latter having opposite sign for low applied magnetic fields. This indicates the presence of an irreversible contribution to the magnetization of $CaFe_2O4$, which can not be switched below a critical field. Moreover, the presence of a spontaneous magnetization is supported by hysteresis (Fig. \ref{fig:5}c) in the $M-H$ loops and their vertical shift, the latter appearing when the sample is cooled in a magnetic field (Fig. \ref{fig:5}d). Vertical shifts in the $M-H$ loops under field-cooling have been observed before in uncompensated antiferromagnets\cite{Henne2016Exchange-bias-likeAntiferromagnet} or inhomogeneous systems characterized by ferrimagnetic moments embedded in a AF matrix. The local magnetic response of $CaFe_2O_4$ films, studied by means of low temperature MFM (Fig. \ref{fig:8}), is also consistent with the presence of a magnetic moment: the MFM magnetic signal, which is only sensitive to out-of-plane magnetization, appears below 185 K, and is often localized at the borders of the domains or needle bunches. The observed contrast is opposite (field in- and field out- of the plane) at both sides of the needles, in good agreement with the expected behaviour of magnetic moment aligned along the needle long axis direction, which produces magnetic field lines that bend in the out-of-plane direction when the needles end.

Thus, the overall magnetic response of $CaFe_2O_4$ thin films is more consistent with an uncompensated AF behaviour than pure AF behaviour. In oxygen-deficient polycrystalline samples, Das and coworkers \cite{Das2016Self-AdjustedCaFe2O4} detect the formation of ferrimagnetic clusters induced by oxygen vacancies. These accumulate at the domain boundaries and, by requirement of charge neutrality, introduce a proportional amount of $Fe^{2+}$ which in turn causes incomplete cancellation of the magnetic moments. Oxygen vacancies are also common in oxide thin films grown by means of PLD. Thus, it is possible that oxygen vacancies are also present in our films, despite having annealed them in 200 mbar oxygen atmosphere after the growth. However, the absence of $Fe^{2+}$ signature in Mössbauer Spectrometry experiments (Fig. \ref{fig:6}) suggests that the spontaneous magnetization of our samples does not originate from oxygen-vacancies induced ferrimagnetic clusters. More consistently with our data, the net magnetization in $CaFe_2O_4$ can be caused by the formation of ``orphan spins" at the boundaries between different magnetic \cite{Stock2017OrphanCaFe2O4}. This scenario seems supported by the fact that the largest $M-H$ hysteresis is observed at 130 K (Fig. \ref{fig:5}c), where the coexistence of A and B phases is expected to be maximum.

Another possibility is that the net magnetic moment arises from locally uncompensated moments at the boundaries between domains. This would be in line with the scenario recently suggested by Songvilay \textit{et al.}\cite{Songvilay2020Disorderx=00.5}, which does not require the coexistence of two phases with long-range order in a wide temperature range. The presence of crystallographic domains, as those detected in our films, which provide ``fixed'' magnetic domain walls below the transition temperature, could also play an important role in this scenario. The behavior of the order parameter, extrapolated from Mössbauer Spectrometry data in our samples reveals critical behavior at the Nèel transition of the A phase: this supports the existence of strong fluctuations of the staggered magnetization at T$_N$, decaying away from the transition, interestingly similar to the behaviour of the B-phase order parameter in refs.[\cite{Corliss1967MagneticFerrite,Songvilay2020Disorderx=00.5}]. At the same time, the critical behaviour of the transition contradicts the hypothesis of nucleation of the A phase at the boundaries of the B phase as the transition mechanism, as that would give rise to a discontinuous phase transition.

Previous studies, reported a broad feature in the $\chi$ vs $T$ plot above $T_N$ \cite{Gandhi2017MagnetocrystallineCrystal, Das2018FerrimagneticCaFe2O4delta}, that can be fitted using the Bonner-Fisher model for linear magnetic chains with anisotropic coupling\cite{Bonner1964LinearCoupling}. This might indicate the existence of short-range and low-dimensional AF exchange, before reaching three-dimensional long-range ordering. However, the absence of hyperfine magnetic splitting at room-temperature in Mössbauer Spectrometry experiments contradicts the hypothesis of room-temperature interaction between $Fe^{3+}$ spins in the samples of this study. \\

To conclude, $CaFe_2O_4$ thin films have been grown for the first time on $TiO_2$ (110) substrates by means of PLD with thickness in the order of 100 nm. The films form crystal domains that consist of needle-like crystals with the long axis along the magnetic easy axis, displaying a clear epitaxial relation with the substrate. The magnetic properties of the $CaFe_2O_4$ thin films studied by means of SQUID magnetometry, Mössbauer spectrometry and low-temperature MFM are consistent and reveal an ordering temperature of about 185 K, concomitant with the presence of a net magnetic moment along the b-axis. The vertical shifts of the $M-H$ loops depending on the field-cooling conditions, evidence that the observed net magnetic moment is not standard ferrimagnetic behaviour. The results are consistent with an antiferromagnet with uncompensated moments but the role played by the crystallographic and/or magnetic domains needs to be clarified. A single A-phase ground state is detected and the critical nature of the transition is revealed with a $\beta$ exponent consistent with the 3D Ising antiferromagnet universality class, precluding nucleation and growth as a plausible mechanism for the transition.\\

Outlook: Further characterization of the magnetic structure of $CaFe_2O_4$ films is needed to completely explain our results. Important questions are still open regarding the stability and coexistence between the A and B magnetic phases observed in bulk samples, the role of critical fluctuations in the stabilization of the B phase and the influence of epitaxial strain on the magnetic phase diagram. Eventually, our goal is to control the relative stability of the A and B phases, in order to obtain a highly responsive system at the boundary between multiple spatial modulations. We believe that $CaFe_2O_4$ thin films represent an interesting prospective system for the study of ``spatial chaos''\cite{Jensen1985SpatialChaos} arising from competing interactions. In such systems, the presence of multiple accessible states close in energy, leads to enhanced susceptibility and adaptability, that are crucial for applications in adaptable electronics, such as neuromorphic computing. Finally, the polar nature of the domain boundaries of the $CaTiO_3$ layer provides an opportunity to explore the multiferroic properties of these $CaTiO_3$/$CaFe_2O_4$ self-organized heterostructures. 

%%%%%%%%%%%%%%%%%%%%%%%%%%%%%%%%%%%%%%%%%%%%%%%%%%%%%%%%%%%%%%%%&&
\section*{Methods}
Sample growth\\The $CaFe_2O_4$ films of this study have been deposited by PLD using a KrF ($\lambda$=248 nm) excimer laser. The target was a home-made ceramic pellet of $CaFe_2O_4$, prepared by solid state synthesis \cite{PHILLIPS1958PhasePressure, Yin2013InfluenceState, Rao1992ChemicalMaterials} from $CaCO_3$ (3N Sigma Aldrich) and $Fe_2O_3$ (99.998$\%$ Alfa Aesar) precursors. The powders were mixed and milled in an agate ball mill at 200 rpm for 2 hours and pressed into a 20 mm diameter pellet with 9.5 tons. Calcination and sintering were executed at 600 \si{\celsius} and 1200 \si{\celsius} respectively. The crystal structure was determined to be single phase $CaFe_2O_4$ via XRD using a Panalytical X'Pert Pro diffractometer in Bragg Brentano geometry. Prior to growth, single crystal $TiO_2$ (110) substrates (CrysTec Gmbh) were treated to reveal the step edges \cite{Yamamoto2005PreparationProperty,Ahmed2014WetConditions} by etching for 1 min with buffered oxide etch (BHF) followed by 1 hour annealing at 900 \si{\celsius} under a constant oxygen flux of 17 l/h.
The optimal growth parameters were determined to be as follows. The laser was focused on the target positioned at 50 mm from the substrate with a spot size of  1.8 mm\textsuperscript{2} . The laser fluence and frequency were 2.8 J/cm\textsuperscript{2} and 10 Hz respectively. The substrate temperature during growth was 850 \si{\celsius} and the  partial oxygen pressure ($P_{O2}$) in the chamber 0.2 mbar.  After deposition the samples were cooled with a rate of -1 $^\circ$/min in $P_{O2}$=200 mbar. The number of pulses was varied in a range from 6000 to 15000 to obtain different film thicknesses. The film surface was monitored during growth via in-situ RHEED. \\
\vskip 0.1in
\noindent
Structural characterization\\Characterization of the films surface was performed using AFM (Bruker Dimension XR microscope) and SEM (FEI Nova NanoSEM 650). XRD measurements were done with a laboratory diffractometer (Panalytical X’Pert MRD Cradle), using Cu K$\alpha$ radiation (1.540598 $\si{\angstrom}$). TEM experiments were conducted on a Cs corrected Themis Z (Thermofischer inc.) microscope. Electron beam was operated at a high tension of 300 kV, and STEM imaging was performed at a beam convergence angle of 23.5 m rad. HAADF-STEM images were acquired with an annular detector in the collection range of 65-200 mrad. DPC images were obtained and analysed using segmented detectors. EDS spectra were collected in the ChemiSTEM mode with 4 symmetric detectors along the optical axis. \\
\vskip 0.1in
\noindent
Mössbauer Spectrometry\\The samples used for Mössbauer Spectrometry were grown from a \textsuperscript{57}$Fe$ enriched target with the same parameters as above. The target was synthesized as described before, but adding to the standard $Fe_2O_3$ precursors $80\%$ of the enriched oxide, prepared by annealing of \textsuperscript{57}$Fe$ powders at 800 \si{\celsius} for 2 hours in a constant oxygen flow of 18 l/h \cite{Lysenko2014TheAnalysis}. CEMS measurements were performed in normal incidence using a home-made gas flow ($He-CH_4$) proportional counter \cite{Juraszek2009AFilms}. For the measurements at low temperatures, the counter was mounted inside a closed-cycle He cryostat \cite{Sougrati2012GasExperiments}. The source was $\textsuperscript{57}Co$ in Rh matrix of about 1.85 GBq activity, mounted in a velocity transducer operating in constant acceleration mode. The spectra were least squares fitted using the histogram method and assuming Lorentzian lines. Isomer shifts are given with respect to $\alpha-Fe$ at 300 K.
\vskip 0.1in
\noindent
Magnetometry and data analysis\\The magnetic properties were studied by means of SQUID magnetometry (Quantum Design MPMS-XL 7) with RSO option in a range of temperature varying from 5 K to 400 K and at fields ranging from 100 Oe up to 7 T. The field was applied either parallel or perpendicular to the magnetization direction of the structural domain with [010] parallel to the substrate [1-10] direction. The long moment values obtained from the SQUID-MPMS has been analyzed using Origin software as follows. First the experimental data has been subtracted of the signal of a clean substrate, measured in the same conditions as the sample. This introduces a small error due to the fact that in the data used as background reference does not contain the signal of the intermediate $CaTiO_3$ layer formed during growth. Then, the experimental data (in emu) has been divided by the magnetic field (in Oe) and the number of moles to yield the magnetic susceptibility of $CaFe_2O_4$ in emu/mol Oe (for the $M-H$ loops, the magnetization has been further converted into units of Bohr Magnetrons per formula unit). This step also introduces an error in our estimation, due to the imprecise estimation of the film thickness via TEM, which is necessary to normalized for the amount of material. Therefore, in this study we do not attempt to provide a precise quantitative analysis of the magnetic response. 
\vskip 0.1in
\noindent
Scanning SQUID microscopy\\The experiments were performed with a scanning SQUID microscope \cite{Kirtley1995High-resolutionMicroscope} with a spatial resolution of approximately 5 $\mu$m \cite{Reith2017AnalysingMicroscopy} and field resolution of approximately 50 nT. The samples were cooled and measured in zero background field at 4 K.  Various sets of 12 scans of 250 $\mu$m x 250 $\mu$m size, with 250 $\mu$m spacing in between (total covered area about 1.75 mm x 1.75 mm), were collected in different areas to test for homogeneity of the samples.  
\vskip 0.1in
\noindent
Magnetic force microscopy\\The MFM experiments presented in this study are performed with a customized Attocube scanning probe microscope inserted in a Quantum Design Physical Property Measurement System (PPMS). Multiple scans were collected at different temperatures upon cooling the sample from 300 K to 12 K. In some cases, a magnetic field ranging from -0.1 to 0.1 T was also applied perpendicular to the film surface. The sample surface was scanned using commercial (Nanoworld) $Co-Cr$ coated tips that were magnetized prior to use. The images were collected in dual-pass tapping mode, with a second scan lift of 30 or 50 nm. The data were then processed with the open source software Gwyddion.
%%%%%%%%%%%%%%%%%%%%%%%%%%%%%%%%%%%%%%%%%%%%%%%%%%%%%%%%%%%%%%%%%%%%%%%%%%%%%%%%%%%%%%%%%%
\section*{Data availability}
The datasets generated during and/or analysed during the current study are available from the corresponding author on reasonable request.
%%%%%%%%%%%%%%%%%%%%%%%%%%%%%%%%%%%%%%%%%%%%%%%%%%%%%%%%%
%\bibliography{references2}
%\bibliographystyle{naturemag}

%%%%%%%%%%%%%%%%%%%%%%%%%%%%%%%%%%%%%%%%%%%%%%%%%%%%%%
\section*{Acknowledgements}
The authors are grateful to Maxim Mostovoy for introducing them to this interesting material and to Maria Azhar and Maxim Mostovoy for their insight on the interpretation of the magnetic data. We acknowledge useful scientific discussions with Urs Staub and Hiroki Ueda and Kohei Yoshimatsu. We also gratefully acknowledge the technical support of Jacob Baas, ir. Henk Bonder and ir. dr. Václav Ocelík in performing the experiments of this study. Financial support by the Groningen Cognitive Systems and Materials Center (CogniGron) and the Ubbo Emmius Foundation of the University of Groningen is gratefully acknowledged. P.N. acknowledges the funding received from European Union’s Horizon 2020 research and innovation program under Marie Sklodowska-Curie grant agreement No: 794954 (Project name: FERHAZ) and J.J. acknowledges support from Region of Normandy and the European Regional Development Fund of Normandy (ERDF) through the MAGMA project.
%%%%%%%%%%%%%%%%%%%%%%%%%%%%%%%%%%%%%%%%%%%%%%%%%%%%
\section*{Author contributions statement}
B.N.conceived the project. S.D. designed the experiments, synthesized the samples, performed the basic structural and magnetic characterization and data analysis. P.N. performed the TEM experiments and analyzed the data. J.J. performed the Mössbauer Spectrometry experiments and analyzed the data. P.R performed the scanning SQUID microscopy experiments under the supervision of H.H. S.D., B.N. and P.N. discussed the results. S.D. wrote the manuscript which was reviewed by all the authors. 
%%%%%%%%%%%%%%%%%%%%%%%%%%%%%%%%%%%%%%%%%%%%%%%%%%%
\section*{Competing interests} 
The authors declare no competing interests.

\end{document}

\maketitle
%%%%%%%%%%%%%%%%%%%%%%%%%%%%%%%%%%%%%%%%%%%%%%%%%%%%%%%%%%%%%%

\section{\label{sec:SI1} Target ablation}
\begin{figure}[htb]
\includegraphics[width=\linewidth]{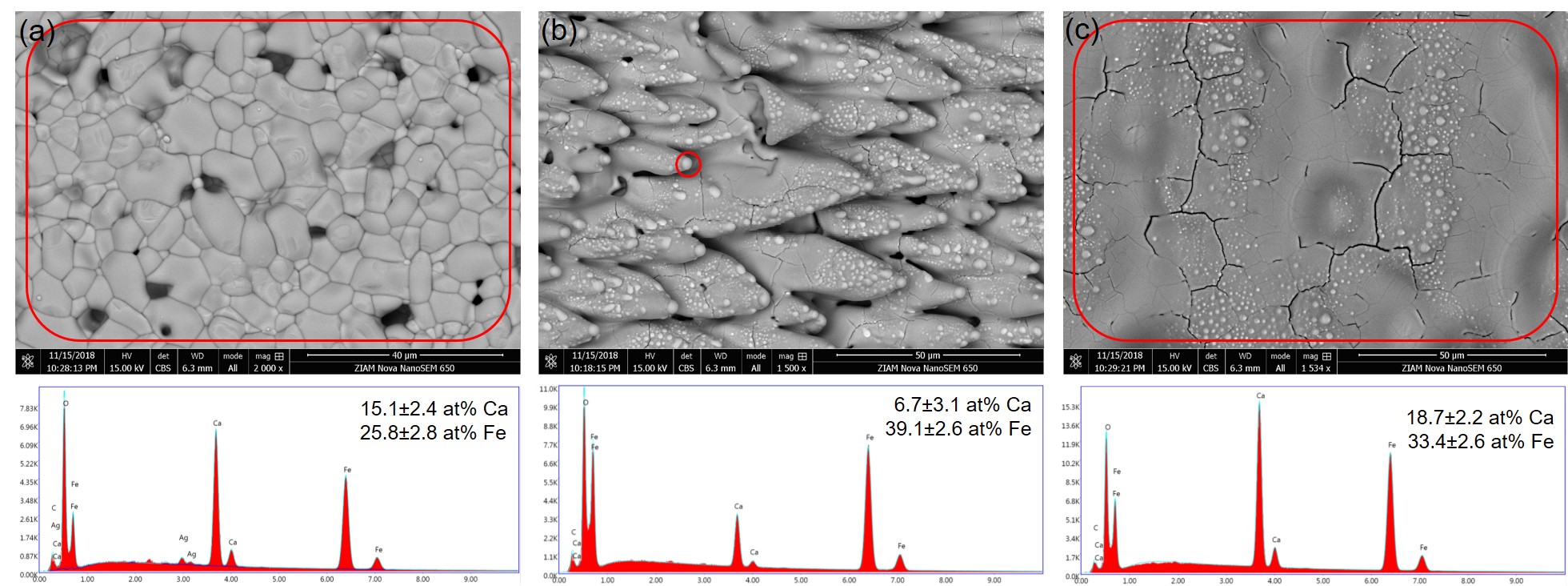}%
\caption{\label{fig:SI1}\textbf{Target ablation.} SEM images collected with a back-scattered electron detector of the ceramic $CaFe_2O_4$ target. \textbf{a} Pristine state, \textbf{b} after ablation at 1 $J/cm^2$ and \textbf{c} after ablation at 2.5 $J/cm^2$. The red line indicates the area from which the under-laying EDS spectrum was collected.}
\end{figure}

For the deposition of $CaFe_2O_4$ films a home-made ceramic pellet of single phase $CaFe_2O_4$ was used as a target. After ablation no change in the XRD peaks position was observed, but only a broadening due to melting of the material. Supplementary Figure \ref{fig:SI1}a shows a SEM image of the as-synthesized target and relative EDS elemental analysis.\\
During the growth optimization process we observed that in order to ablate in equal amount $Ca$ and $Fe$ atoms from the target an high laser fluence (about 2.8 $J/cm^2$) was required. At lower energies, the ablated region of the target under an SEM microscope displays a rough morphology characterized by high pillars of non-ablated material. On top of each pillar an island of $Fe$-rich material is found (lighter contrast in Supplementary Figure \ref{fig:SI1}b). This is due to the higher melting point that prevents $Fe$ to be transferred to the plasma plume thus blocking the laser to reach the material underneath. Therefore, films grown in this condition are $Fe$-deficient. On the other hand, when the laser fluence is above 2.5 $J/cm^2$, the ablated area appears more homogeneous and no difference in composition before and after the deposition is measured via Energy dispersive X-ray spectroscopy (EDS) (see Supplementary Figure \ref{fig:SI1}c). 

\section{\label{sec:SI2} Similarity between (004) and (302) orientations}
In this study we encountered difficulties in the films characterization due to the similarities between $CaFe_2O_4$ (004) and (302) planes. The arrangement of the atoms in such planes is indeed very similar and based on alternating ($Fe-Fe-Ca-Fe-Fe-Ca$) chains of cations ordered along the b-axis, as shown in Supplementary Figure \ref{fig:SI2}a-b. This results in almost indistinguishable, besides the modulation of the spots intensity, electron diffraction patterns (see Supplementary Figure \ref{fig:SI2}c). 
\begin{figure}[htb]
\includegraphics[width=\linewidth]{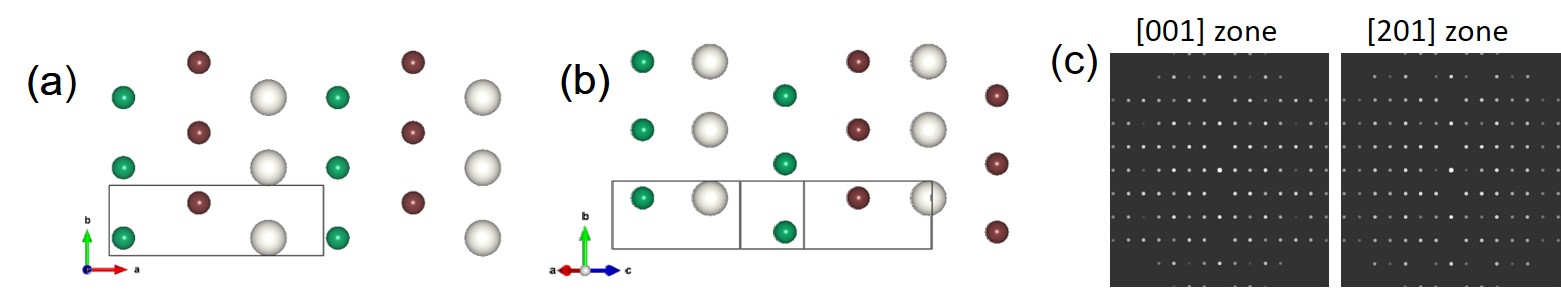}%
\caption{\label{fig:SI2} \textbf{Similarity between (004) and (302) orientations.} Schematic representation of cations arrangement in $CaFe_2O_4$ planes. $Ca$, $Fe(1)$, and $Fe(2)$ atoms are depicted in white, green and brown respectively. \textbf{a} $CaFe_2O_4$ 004 plane \textbf{b} $CaFe_2O_4$ (302) plane \textbf{c} simulation of electron diffraction pattern from the (001) and (201) zones. }
\end{figure}

\section{\label{sec:SI3} EDS}
\begin{figure}[htb]
\begin{centering}
\includegraphics[width=.8\linewidth]{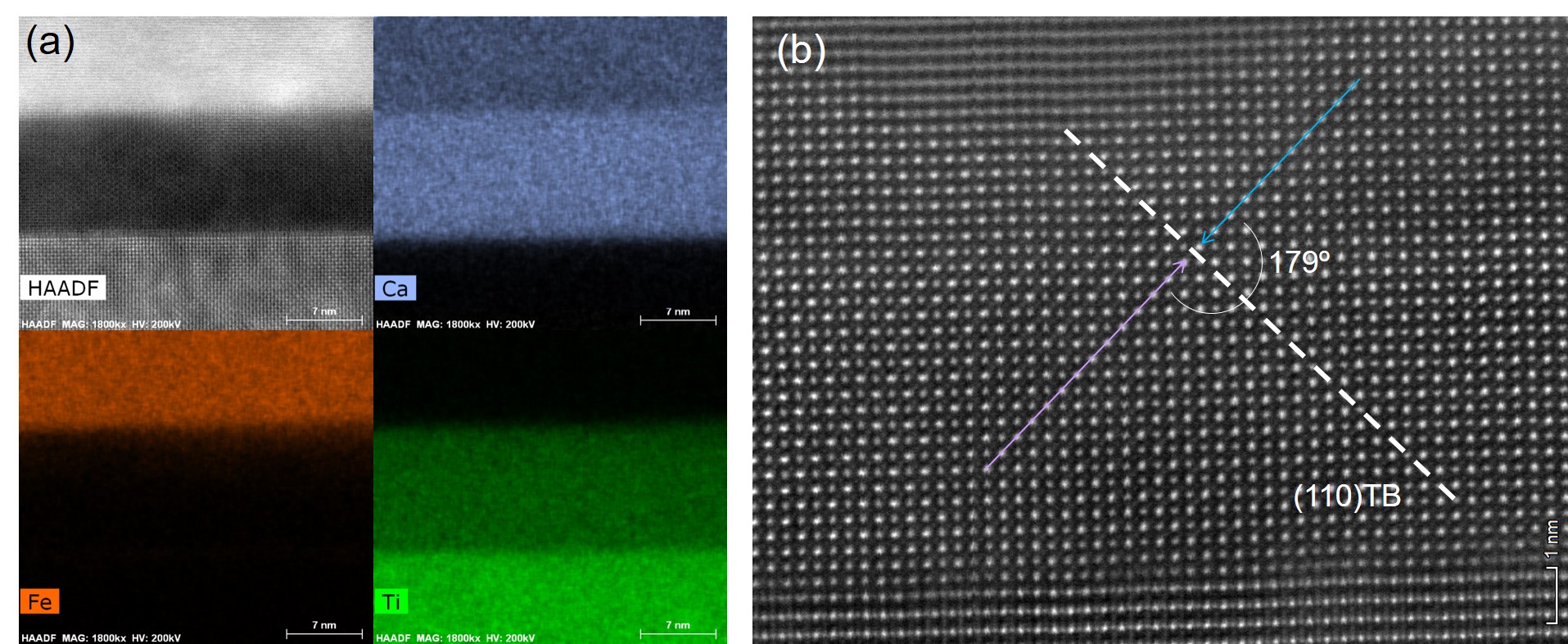}%
\caption{\label{fig:SI3} \textbf{Chemical analysis of the thin-film stack and domain boundaries in CaTiO\textsubscript{3} layer.} \textbf{a} Elemental distribution of $Ca$ blue, $Fe$ orange and $Ti$ green across the three layers obtained through STEM-EDS, clearly revealing a ~10 nm Calcium Titanate layer at the interface. \textbf{b} HAADF-STEM image of the $CaTiO_3$ layer imaged along the [001] zone axis where a (110) ferroelastic domain wall (white-dashed line) can be seen. The purple and blue arrows on either side of the domain wall show an orientation mismatch of about 1$^\circ$ between the domains, corresponding to the well-reported 179$^\circ$ domains in bulk $CaTiO_3$}
\end{centering}
\end{figure}

Supplementary Figure \ref{fig:SI3}a shows the distribution of $Ca$, $Fe$ and $Ti$ atoms through a cross section of our films measured by means of energy dispersive X-ray spectroscopy (EDS). $Fe$ and $Ti$ contrast is only observed in the upper and lower layers respectively while $Ca$ is found everywhere. Combination of EDS with imaging of the lattice by means of TEM allowed us to determine that a perovskite $CaTiO_3$ (10 nm) layer is present at the interface between $CaFe_2O_4$ film and $TiO_2$ substrate. Based on the oxygen tilts, and the lattice parameters, we can deduce that the $CaTiO_3$ layer crystallizes in an orthorhombic structure with \textit{Pnma} space group. It is oriented along the b-axis out of plane, and contains ferroelastic domains separated by polar (110) domain walls (Supplementary Figure \ref{fig:SI3}b), reminiscent of bulk $CaTiO_3$. Possible magnetoelectric coupling between these domain walls and the $CaFe_2O_4$ layer is a subject of future investigation. \\

\section{\label{sec:SI4} EBSD}
\begin{figure}[htb]
\begin{centering}
\includegraphics[width=.45\linewidth]{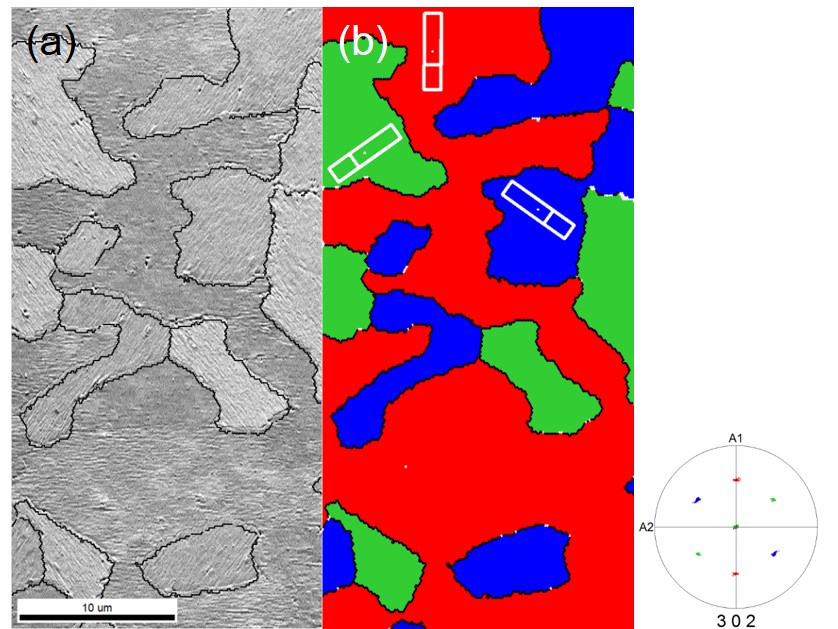}%
\caption{\label{fig:SI4}\textbf{EBSD measurement.} \textbf{a} SEM image (collected with a secondary electron detector) and grain boundaries (>5$^\circ$). \textbf{b} Orientation map and pole figure of a 96 nm thick film. Red, green and blue indicate crystals with the same out-of-plane orientation, namely [302], but different in-plane orientation. A pole figure of the orientations is shown in the inset.}
\end{centering}
\end{figure}

Supplementary Figure \ref{fig:SI4} shows the results of orientation imaging microscopy performed of $CaFe_2O_4$ thin films. The experiments are performed by means of Electron backscatter diffraction (EBSD) in a SEM. The color map and corresponding pole figure of \ref{fig:SI4}b show the orientation of the domains imaged in \ref{fig:SI4}a. Three crystal orientations are found, all with [302] out-of-plane but different in-plane directions. In the map the red color indicates areas of the film where the $CaFe_2O_4$ [010] is parallel to the substrate [1-10], while red and blue indicate domains with a $\pm$55$^\circ$ tilt from it. These results are in good agreement with what found in x-ray pole figures and RSMs shown in the main text.

\section{\label{sec:SI5} Bulk magnetic properties}
The magnetic response of the $CaFe_2O_4$ ceramic target used for the films deposition has been studied by means of SQUID magnetometry. Plots of the susceptibility as a function of temperature are reported in Supplementary Figure \ref{fig:SI5}a and b measured in a 100 Oe and 2 T fields respectively. The response is consistent with the literature data regarding bulk $CaFe_2O_4$. with a Néel temperature of 175 K and a strong ZFC/FC splitting in the low field measurement.
\begin{figure}[htb]
\begin{centering}
\includegraphics[width=\linewidth]{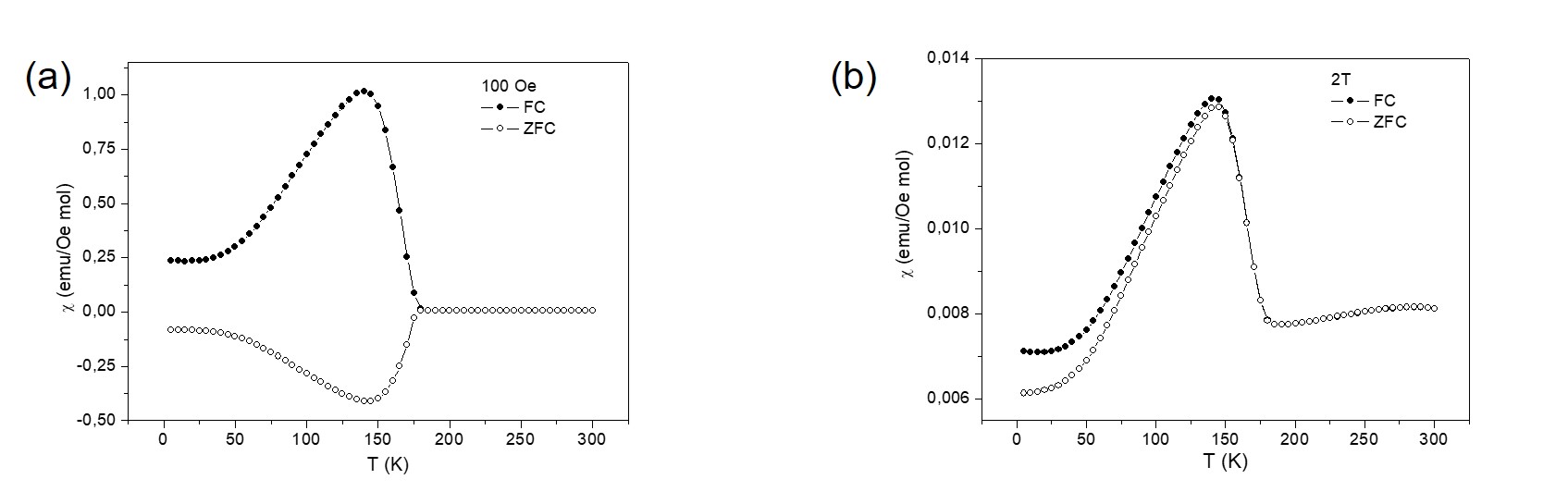}%
\caption{\label{fig:SI5}\textbf{Bulk magnetic properties.} Plot of the magnetic susceptibility ($\chi$) as a function of temperature ($T$) from 5 to 300K in a \textbf{a} 100 Oe and \textbf{b} 2 T applied field.}
\end{centering}
\end{figure}

\section{\label{sec:SI6}Macro scale magnetic properties}
\begin{figure}[htb]
\begin{centering}
\includegraphics[width=\linewidth]{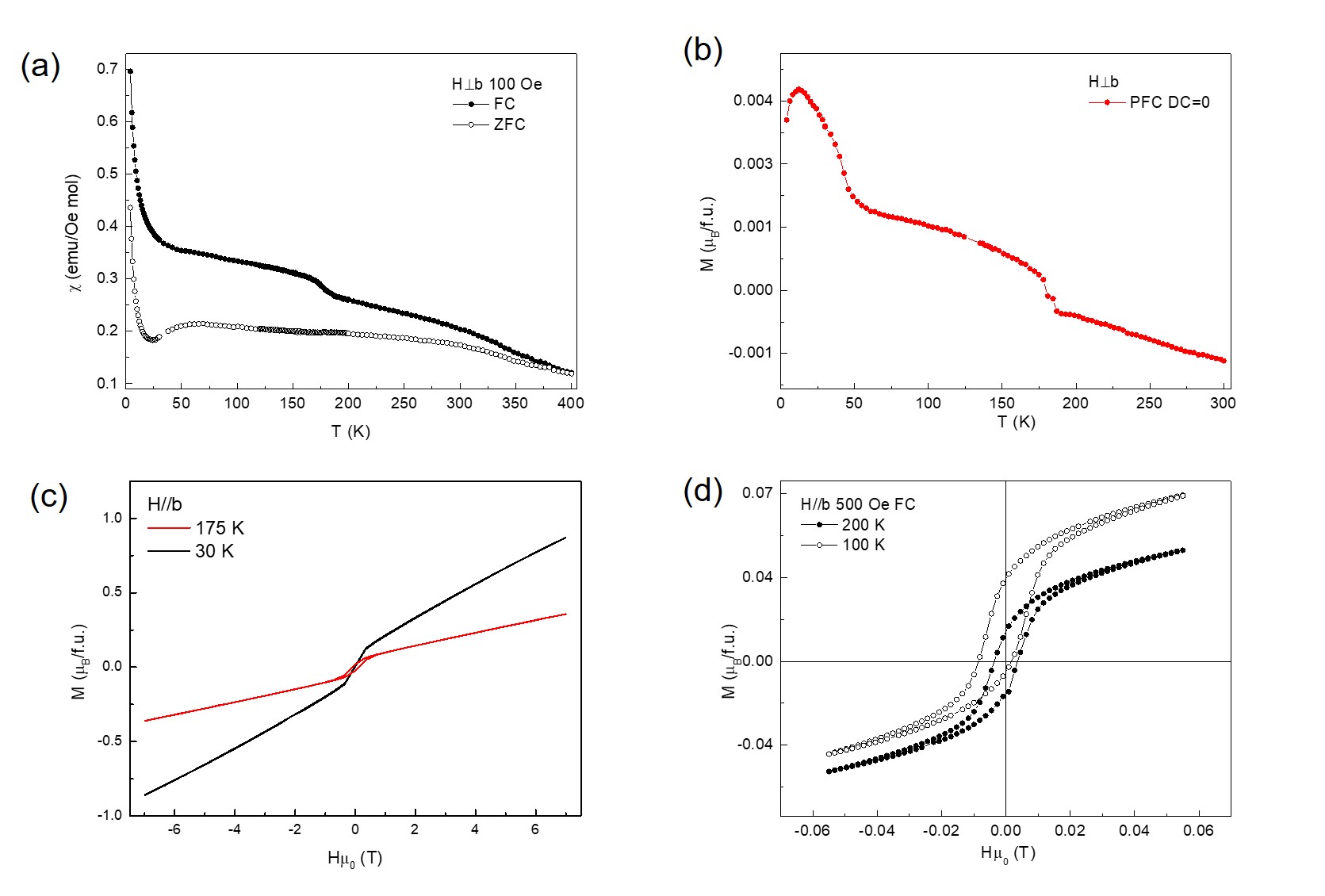}%
\caption{\label{fig:SI6} \textbf{Additional SQUID magnetometry measurements}\textbf{a} Plot of the magnetic susceptibility ($\chi$) of a 84 nm thick sample as a function of temperature ($T$) in a 100 Oe magnetic field perpendicular to the b-axis. \textbf{b} Magnetization ($M$) of a 90 nm thick sample measured as a function of temperature in zero applied DC field after cooling in 100 Oe field applied perpendicular to the b-axis. \textbf{c} Plot of the $M-H$ loops measured at 175 (red) and 30 (black) K from 7 to -7 T, with $H$ parallel to the b-axis. \textbf{d} Plot of the $M-H$ loops measured at 200 (full dots) and 100 (open dots) K from 550 to -550 Oe, with $H$ parallel to the b-axis.}
\end{centering}
\end{figure}

To complement the SQUID measurement data shown in the main text, Supplementary Figure \ref{fig:SI6}a and b display the response of of $CaFe_2O_4$ thin films to fields applied perpendicular to the magnetization direction. In Supplementary Figure \ref{fig:SI6}a the FC and ZFC susceptibility ($\chi$) is plotted as a function of temperature at 100 Oe. The same features as for the parallel measurement are found (paramagnetic tail at low-$T$, $T_N$=188 K, ZFC/FC splitting), albeit lower values are measured. In particular, the maximum reached by $\chi$ below $T_N$ is almost absent for $B\perp$ to b. Moreover, the ZFC curve does not show sign reversal. Supplementary Figure \ref{fig:SI6}b shows the magnetization measured at zero applied DC field after positive field cooling. Here again an analogous trend is observed, except for the absence of the paramamgnetic contribution below 20 K.\\ 
$M-H$ loops were also collected at different temperatures (5-30-100-130-175-200-300 K) to study the evolution of the spontaneous magnetization with the ordering of A and B magnetic phases. The same results are obtained upon directly measuring the loops during cooling or by heating the sample above $T_N$ between each loop. Fig. 5c of the main text shows the loop at 130 K, where the maximum hysteresis appears. Here (Supplementary Figure \ref{fig:SI6}c) we report the $M-H$ loops collected at 175 and 30 K for comparison. At these temperatures lower and no magnetic hysteresis is observed, respectively, consistently with the disappearance of antiphase boundaries. \\
Supplementary Figure \ref{fig:SI6}d displays the absence of field-cooling induced vertical shift in the $M-H$ loops, when the measurement is performed above $T_N$, despite a small hysteresis persists up to 300 K.\\

\section{\label{sec:SI7}Low temperature MFM}
\begin{figure}[htb]
\begin{centering}
\includegraphics[width=\linewidth]{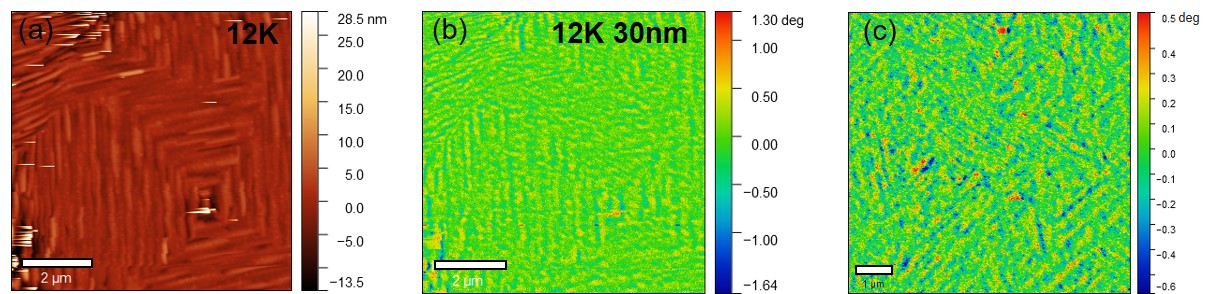}%
\caption{\label{fig:SI7} \textbf{MFM at 12K in Attocube AFM (PPMS insert) of a 120 nm thick CaFe\textsubscript{2}O\textsubscript{4} film.} \textbf{a} Topography and \textbf{b} dual-pass phase recorded using a non magnetic tip. \textbf{c} Difference between Fig. 8h and Fig. 8i of the main text (dual-pass phase at 0.05 and -0.1 T applied field respectively), calculated with the software gwyddion.}
\end{centering}
\end{figure}

To further prove that the signal observed in the phase of the low-temperature MFM experiments (Fig. 8 of the main text) is of magnetic nature, we repeated the measurement using a non magnetic tip. The experiment is performed in dual-pass mode: first the topography is recorded in tapping mode, then the tip is lifted of a fixed height above the sample surface and the phase of the second-pass signal is monitored. Supplementary Figure \ref{fig:SI7}a and b show the topography and MFM phase recorded with a non magnetic tip. Here, the strong contrast in the phase signal is absent, even at the lower lift height of 30 nm. In Supplementary Figure \ref{fig:SI7}b only some residual cross-talk with the surface topography can be seen, which is also visible in Fig. 8b of the main text at 30 nm lift height.\\
Supplementary Figure \ref{fig:SI7}c shows the difference between Fig. 8h and Fig. 8i of the main text (recorded under applied fields of opposite signs). It can be seen that the majority of the signal cancels out, except for the cross-talk with the image topography that is enhanced. This is because, at opposite field polarities the magnetization of the $Co-Cr$ coated tip is reversed, but not that of the sample, which is perpendicular to the applied magnetic field. Thus the interaction between tip and sample switches from FM to AF, and vice versa, upon reversing field sign.\\
This allows us to conclude that the signal observed in low-T MFM experiment has magnetic origin and can be separated by the topographical information despite the films high surface roughness.\\
\section{Preferential domain}
As discussed in the main text, our samples are constitued by three crystallographic domains. However, across the film the one oriented with the [010] direction parallel to the substrate [1-10] (Domain 1) is preferred. This allows us to align the magnetization axis [010] of most of the sample parallel or perpendicular to the magnetic field $H$ in magnetic measurements.\\
A proof of this is illustrated in Supplementary Figure \ref{fig:S8} that shows three reciprocal space maps around the $CaFe_2O_4$ (600) peak at 2$\theta$=60.25$^\circ$, $\omega$=-1$^\circ$ and $\chi$=0$^\circ$. The first map is collected at $\phi$=0$^\circ$, while the second and third at $\phi$=55$^\circ$ and $\phi$=-55$^\circ$. As it can be seen, the intensity of the peaks is not identical, with the first being higher than the others. 
\begin{figure}[htb]
\centering
\includegraphics[width=\linewidth]{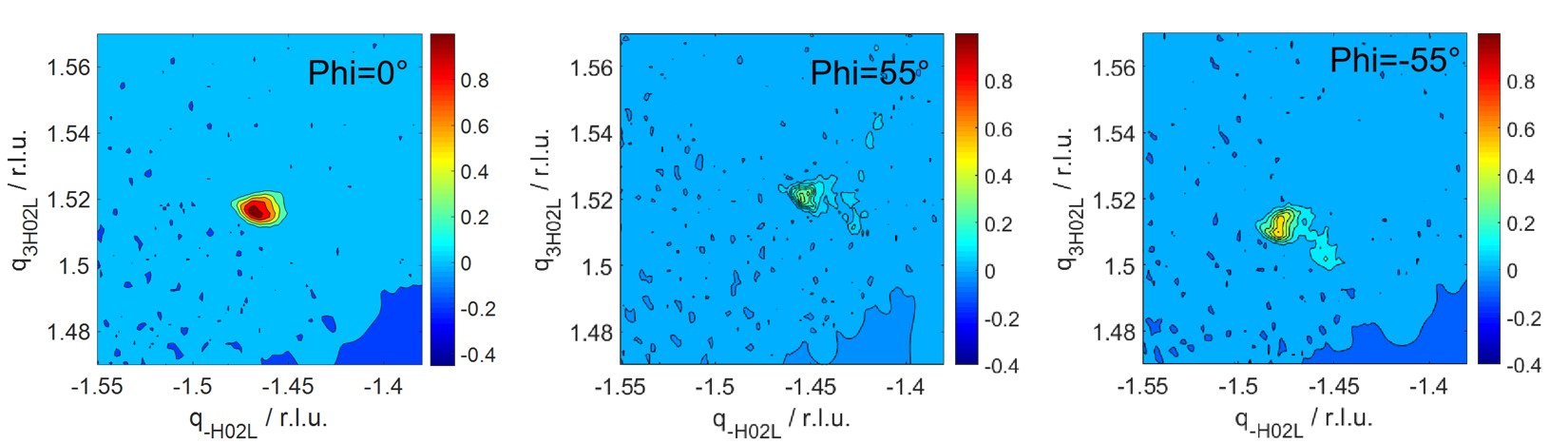}%
\caption{\textbf{Preferential domain.} Reciprocal space maps of the (600) peak collected at \textbf{a} $\phi$=0$^\circ$, \textbf{b} $\phi$=55$^\circ$ and \textbf{c} $\phi$=-55$^\circ$ also confirming the presence of three structural domains, with the first being the principal orientation.}
\label{fig:S8}
\end{figure}